\documentclass{SCIS2025}
\usepackage{diagbox}
\begin{document}

\ArticleType{RESEARCH PAPER}
\Year{2025}
\Month{January}
\Vol{68}
\No{1}
\DOI{}
\ArtNo{}
\ReceiveDate{}
\ReviseDate{}
\AcceptDate{}
\OnlineDate{}
\AuthorMark{}
\AuthorCitation{}

\title{Encrypted-state quantum compilation scheme based on quantum circuit obfuscation for quantum cloud platforms}{}

\author[1]{Chenyi Zhang}{}
\author[1]{Tao Shang}{{shangtao@buaa.edu.cn}}
\author[2]{Xueyi Guo}{}
\author[3]{Yuanjing Zhang}{}

\address[1]{School of Cyber Science and Technology, Beihang University, Beijing 100083, China}
\address[2]{Beijing Academy of Quantum Information Sciences, Beijing 100193, China}
\address[3]{Future Science and Technology Research Lab, China Mobile (Suzhou) Software Technology Company Limited, Suzhou 215163, China}

\abstract{
With the rapid advancement of quantum computing, quantum compilation has become a crucial layer connecting high-level algorithms with physical hardware. In quantum cloud computing, compilation is performed on the cloud platforms, which expose user circuits to potential risks such as structural leakage and output predictability. To address these issues, we propose the encrypted-state quantum compilation scheme based on quantum circuit obfuscation (ECQCO), the first secure compilation scheme tailored for the co-location of compilers and quantum hardware for quantum cloud platforms. It applies quantum homomorphic encryption to conceal output states and instantiates a structure obfuscation mechanism based on quantum indistinguishability obfuscation, effectively protecting both functionality and topology of the circuit. Additionally, an adaptive decoupling obfuscation algorithm is designed to suppress potential idle errors while inserting pulse operations. The proposed scheme achieves information-theoretic security and guarantees computational indistinguishability under the quantum random oracle model. Experimental results on benchmark datasets demonstrate that ECQCO achieves a total variation distance (TVD) of up to 0.7 and a normalized graph edit distance (GED) of 0.88, enhancing compilation-stage security. Moreover, it introduces only a slight increase in circuit depth, while keeping the average fidelity change within 1\%, thus achieving a practical balance between security and efficiency.
}

\maketitle

\section{Introduction}

Quantum computing has experienced rapid development and has demonstrated the potential to outperform classical computers in solving certain complex problems. It is anticipated to drive scientific discoveries in a variety of fields, including cryptography \cite{monz2016realization}, biomedicine \cite{cao2018potential}, and materials science \cite{bauer2020quantum}. However, owing to the expensive cost and maintenance difficulties of quantum computers, users must rely on quantum cloud platforms provided by research institutions and commercial enterprises, such as Origin Quantum \cite{zou2025qpanda3}, IBM Quantum \cite{chow2021ibm}, and Microsoft Azure Quantum \cite{prateek2023quantum}. Through these platforms, users submit their quantum program designs to remote servers, where quantum compilers translate high-level quantum algorithms into executable instructions tailored for specific quantum hardware. The quantum compilation process improves gate-level compatibility, reduces circuit depth and noise sensitivity, and ensures that the resulting quantum circuits can be executed correctly on the target quantum processor.

However, as third-party quantum compilers and quantum hardware deployed in untrusted quantum cloud environments become more widely adopted, the compilation stage of quantum circuits encounters a range of security risks. Adversaries may exploit various vectors, including crosstalk induced by fault injection \cite{ash2020analysis}, insertion or modification of quantum functions via Trojan software \cite{das2023trojannet,das2024trojan,roy2024hardware,john2025quantum}, side-channel leakage through pulse-level power analysis \cite{xu2023exploration,trochatos2023hardware}, and malicious behaviors from untrusted compilers, which can result in cloning, tampering, or reverse engineering of circuit designs \cite{yang2024multi}. Since quantum circuit structures often constitute valuable intellectual assets, it is necessary to protect them against compiler-related threats.

Several existing methods have been proposed to protect quantum circuits \cite{aboy2022mapping}. One approach inserts reversible gates with random parameters into the circuit \cite{suresh2021short,das2023randomized,naz2023reversible}. Another splits a circuit into two or more parts and compiles them separately \cite{saki2021split,upadhyay2022robust,wang2025tetrislock,patel2023toward}. A third adds key qubits that control specific gates within the original structure \cite{topaloglu2023quantum,liu2025loq,rehman2025opaque,raj2025quantum}. Most quantum compilers today are embedded within cloud-based quantum platforms \cite{golec2024quantum}. In contrast, many existing protection models assume that the compiler and the platform are separate. The assumption simplifies circuit recovery after encryption but does not match actual execution workflows in cloud environments. Many of these schemes also lack formal proofs of correctness and do not provide complete security analysis. They often rely on algebraic transformations or circuit structure design to achieve obfuscation. Experimental validation alone cannot answer two key questions: do such obfuscation strategies always work, and what level of security can they provide?

Another research direction in quantum circuit obfuscation is to theoretically explore the extent to which general quantum circuits can achieve a weakened form of black-box obfuscation. Since quantum black-box obfuscation has been proven impossible, researchers have proposed various notions of quantum indistinguishability obfuscation\cite{broadbent2021constructions}, including indistinguishability obfuscation of null quantum circuits\cite{bartusek2021indistinguishability} and quantum-state indistinguishability obfuscation\cite{coladangelo2024use}. In addition, some researchers have suggested universal quantum black-box obfuscation for specific classes of quantum circuits, such as quantum point function obfuscation\cite{shang2019obfuscatability} and nonlinear function obfuscation\cite{pan2023universal}. Such theoretical studies rarely address efficient instantiations at the application level. Moreover, it lacks comprehensive protection for both the structure of quantum circuits and the output information.

To address these limitations, the concept of encrypted-state quantum compilation is introduced. The notion of encrypted-state originates from classical trusted computing, where multiple cryptographic techniques are integrated to ensure that data remains usable but invisible throughout the cloud computing process. By introducing encrypted-state computation into the quantum cloud environment, we propose the encrypted-state quantum compilation scheme based on quantum circuit obfuscation (ECQCO). ECQCO assumes a system model where the compiler and quantum computer reside within the same quantum cloud entity. It decomposes the protection goal into the output obfuscation of quantum circuit and the structure obfuscation of quantum circuit. Scheme correctness and security rely on two quantum cryptographic primitives, namely quantum homomorphic encryption (QHE) and quantum indistinguishability obfuscation (QiO). Our scheme employs quantum cryptographic primitives for efficient instantiation. It integrates techniques, such as probabilistic distribution inference, $T/T^{\dagger}$-gate replacement, and adaptive QiO sequence insertion. Additionally, ECQCO is implemented entirely on the client side. It is orthogonal to existing circuit optimization techniques and remains compatible with any current NISQ-era compiler. To the best of our knowledge, it is the first work to apply quantum cryptography to the protection of quantum circuits systematically. The scheme enhances both security and generality without compromising execution efficiency.

The main contributions of our work are:
\begin{enumerate}
    \item \textbf{The first quantum compilation scheme that protects both the output and the structure of quantum circuits on the classical client side}: Based on quantum homomorphic encryption and an efficient instantiation of quantum indistinguishability obfuscation, the scheme is the first to safeguard both the output and the structural information of quantum circuits, thereby enabling encrypted quantum compilation in quantum cloud platforms. Moreover, we have conducted multiple experiments to prove that the scheme is fully deployed on the classical client side, which means it is independent of the cloud and can seamlessly adapt to the quantum cloud services based on various technical approaches.   
    \item \textbf{A quantum obfuscation algorithm that incorporates dynamic decoupling capability}: While encrypting the circuit, the scheme incorporates dynamic decoupling techniques and proposes the adaptive decoupling obfuscation algorithm to minimize the number of additional quantum gates used. The algorithm applies a periodic series of inversion pulses to reduce the impact of additional gates on compilation and execution performance, maintaining high fidelity.
    \item \textbf{Rigorous formal proofs of correctness and security based on quantum cryptography}: We demonstrate the correctness of the scheme using the quantum one-time pad and the probabilistic testing distinguisher. Furthermore, leveraging the quantum random oracle, we prove that the scheme achieves information-theoretic security for output protection and quantum indistinguishability security for structure protection.

\end{enumerate}

The paper is organized as follows. Section \ref{sec2} provides preliminaries of the two quantum cryptographic primitives involved in ECQCO, namely QHE and QiO. Section \ref{sec3} presents the system model and detailed descriptions of ECQCO, as well as its correctness and security analyses. In Section \ref{sec4}, we demonstrate the obfuscation effectiveness of ECQCO and include the evaluations of correctness, overhead, and fidelity analysis. Finally, Section \ref{sec5} concludes the paper and discusses several open questions.

\section{Preliminaries}\label{sec2}

In this section, we briefly introduce two common quantum cryptographic primitives, namely the QHE scheme based on quantum one-time pad schemes, and the QiO scheme via quantum circuit equivalence. These two primitives form the foundation of ECQCO and serve as the source of its correctness and security guarantees.

\subsection{QHE scheme based on quantum one-time pad}

In 2003, Boykin et al. \cite{boykin2003optimal} introduced a quantum one-time pad (QOTP) using Pauli operators, which enabled quantum cryptographic protocols to achieve information-theoretic security.

\definition[quantum one-time pad]{Let $\sigma$ be the density matrix of a $n$-qubit system, $a,b\in\{0,1\}^n$. The quantum one-time pad encryption and decryption procedures are defined as follows.
\begin{align*}
QEnc_{a,b}:\sigma\rightarrow X^aZ^b\sigma Z^bX^a\\
QDec_{a,b}: X^aZ^b\sigma Z^bX^a\rightarrow\sigma
\end{align*}
}

Due to the indistinguishability property of the QOTP, randomly selected keys encrypt the plaintext quantum state into a maximally mixed state. As a result, an adversary gains no information about either the density matrix $\sigma$ or the key $(a,b)$.

In 2013, Liang et al. \cite{liang2013symmetric} formally defined QHE and proposed the first symmetric QHE scheme based on the QOTP.

\definition[QHE scheme based on quantum one-time pad]\label{def_qhe}
{QHE scheme consists of the following four algorithms.
\begin{enumerate}
    \item Key Generation: Randomly generate an encryption key $ek$.
    \item Encryption: Encrypt a plaintext quantum state $\sigma$ using $ek$, and output the ciphertext state $\rho=Enc(ek,\sigma)$.
    \item Homomorphic Evaluation: Apply a quantum circuit $C_q$ to the ciphertext $\rho$, resulting in a ciphertext computation outcome $Eval^{C_q}(\rho)$.
    \item Decryption: Decrypt the evaluated ciphertext $Eval^{C_q}(\rho)$ using the decryption key $dk$, obtaining the result $\sigma'=Dec(dk,Eval^{C_q}(\rho))$.
    If the scheme is symmetric, then $dk=ek$. Otherwise, the decryption key $dk$ is derived from $ek$ through a key update process.
\end{enumerate}
}

QHE typically requires $\mathcal{F}$-homomorphic.
\definition[$\mathcal{F}$-homomorphi] \label{defin_f}
{Let $\mathcal{F}$ be the set of all quantum circuits. A quantum homomorphic encryption scheme is $\mathcal{F}$-homomorphi if for any quantum circuit $C_q$, there exists a negligible function $negl$ such that for all $\lambda$:
$$
\Delta(\sigma',C_q\sigma)=\Delta(Dec(dk,Eval^{C_q}(\rho)),C_q\sigma)\leq negl(\lambda)
$$
}

\subsection{QiO scheme based on quantum circuit equivalence}

Quantum obfuscation is a powerful tool for achieving functional equivalence. The concept originated from the idea of ``protecting circuit information with qubit'' \cite{aaronson2017ten}, firstly. By analogy with the idea of classical obfuscators, Alagic et al. \cite{alagic2016quantum} formally proposed the definition and impossibility results of quantum obfuscation. Starting from the impossibility results of quantum black-box obfuscation, researchers explore the degree of obfuscation that a certain type of quantum circuits can achieve, including quantum point obfuscation \cite{shang2019obfuscatability,zhang2022instantiation}, quantum power obfuscation \cite{jiang2023quantum}, etc. We are more concerned about quantum indistinguishable obfuscation (QiO), which is a weakening of quantum black-box obfuscation, including zero-circuit quantum indistinguishable obfuscation \cite{bartusek2021indistinguishability}, quantum state indistinguishable obfuscation \cite{bartusek2024quantum}, etc. The reason is that the equivalent quantum implementations can realize the same computational functionality. When two equivalent implementations are given as input, a quantum indistinguishability obfuscator produces outputs that are computationally indistinguishable \cite{coladangelo2024use}.

\definition [QiO based on quantum circuits equivalence] \label{def_qio}
{Let $\{Q_\lambda\}_{\lambda\in \mathbf{N}}$ be a family of quantum implementations for the classical function $f$, and $C$ be a family of quantum circuits. A quantum indistinguishability obfuscator for equivalent quantum circuits is a quantum polynomial-time
($QPT$) algorithm $QiO$ that takes as input a security parameter $1^\lambda$ and a pair of quantum implementations $(\rho,C)\in Q_\lambda$, and outputs a pair of $(\rho',C')$. Additionally, $QiO$ should satisfy the following conditions.
\begin{itemize}
    \item Polynomial expansion: There exists a polynomial function $poly(n)$ such that for all $C\in\mathcal{C}$, $\mathcal{C}$ is a quantum circuit family, the size of the obfuscated circuit $C'$ satisfies $|C'|=poly(|C|)$. It means that the size of the obfuscated circuit $C'$ is polynomially bounded in terms of the size of $C$.
    \item Functional equivalence: For any $C\in\mathcal{C}$, $(\rho',C')\leftarrow QiO(\rho,C)$, $C$ and $C'$ are under $\Delta$subpath equivalence.
    \item Computational indistinguishability: For any $QPT$ distinguisher $D$, there exists a negligible function $negl$ such that for all $\lambda$ and two pairs of quantum implementations $(\rho_1, C_1),(\rho_2, C_2)$ of the same function $f$, the distributions of the obfuscated outputs are computationally indistinguishable.
    \begin{align*}
        |\mathrm{Pr}[D(QiO(1^\lambda,(\rho_1,C_1)\rightarrow(\rho'_1,C'_1))=1]-\mathrm{Pr}[D(QiO(1^\lambda,(\rho_2,C_2)\rightarrow(\rho'_2,C'_2))=1]|\leq negl(\lambda)
    \end{align*}
\end{itemize}
}

The equivalence testing of quantum implementations for a classical function $f$ reduces to indistinguishability analogous to quantum states represented by density operators. The simplification relies on applying a constructed unitary transformation to evolve all possible inputs one by one. The approach reflects an implicit strategy commonly adopted in security proofs for general indistinguishability obfuscation schemes. The method becomes increasingly complex as the size of the unitary matrix grows exponentially with the number of qubits \cite{bernstein1993quantum}. It also leads to inherent security degradation in all known indistinguishability obfuscation constructions \cite{jain2022indistinguishability}.

\section{Encrypted-state quantum compilation scheme}\label{sec3}
\subsection{System model}

The quantum circuit compilation scenario involves a trusted client and an untrusted server. The client submits a quantum program to the server, where the quantum program is represented as a quantum circuit. The server performs quantum compilation, execution, and measurement, and returns the result to the client. To ensure the soundness and robustness, we establish the following assumptions.
\begin{itemize}
    \item The client does not possess quantum computational capabilities.
    \item The server is assumed to be a passive adversary that eavesdrops during the three phases described above.
\end{itemize}

Given that the server is semi-honest, two types of security threats arise during the quantum compilation phase.
\begin{itemize} 
    \item Leakage of output information: The server obtains the result of the quantum program after execution and measurement on the quantum hardware. Since the quantum state carries information about the quantum circuit, and the client cannot process quantum data, the server has access to both the input and output quantum states. It allows the server to effectively reconstruct the entire quantum program. 
    \item Leakage of structural information: The server gains knowledge of the structure of the submitted quantum circuit, including its topology (as a directed acyclic graph), the number and types of quantum gates, and the circuit depth. Such information can reveal sensitive intellectual property of the client.
\end{itemize}

Quantum compilation typically alters the circuit structure significantly, while the output quantum state remains unmodified on the server side. Therefore, eavesdropping is effective only during the quantum compilation phase, which justifies our design goal of achieving encrypted-state quantum compilation.

\begin{figure}
    \centering
    \includegraphics[width=0.8\linewidth]{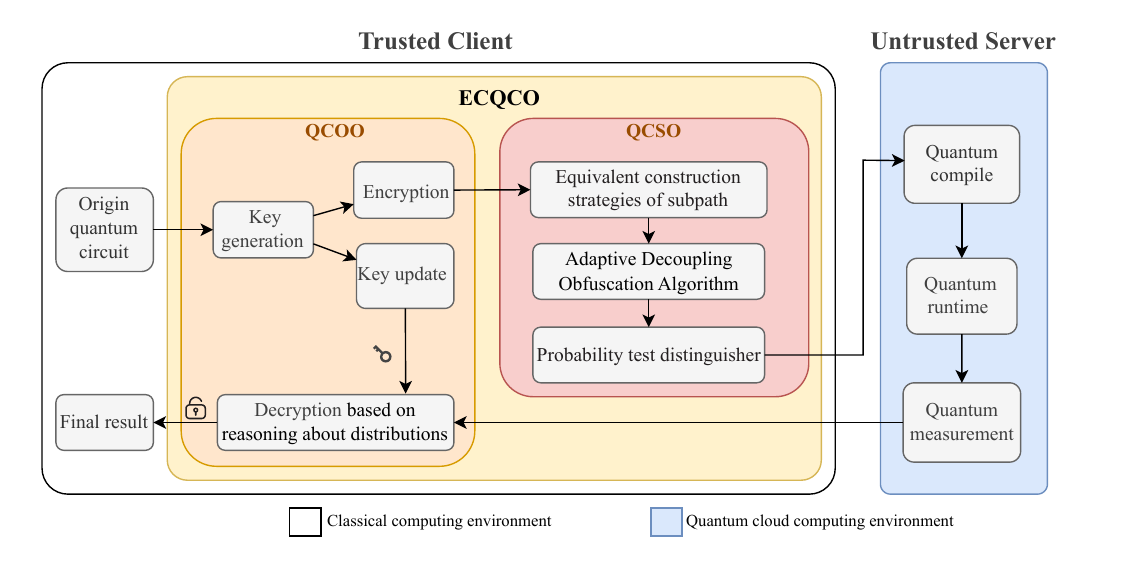}
    \caption{ECQCO scheme}
    \label{ecqco}
\end{figure}

We will propose the encrypted-state quantum compilation scheme based on quantum circuit obfuscation (ECQCO), which aims to address security threats arising during the quantum compilation phase. As illustrated in Figure \ref{ecqco}, ECQCO consists of two core components, quantum circuit output obfuscation (QCOO) and quantum circuit structure obfuscation (QCSO), which mitigate the risks of output information leakage and structural information leakage, respectively.

ECQCO is executed entirely on the client side. The client firstly applies QCOO and QCSO in sequence to encrypt and obfuscate the designed quantum circuit. Then the client submits the protected circuit to the server. Upon receiving the execution result from the server, the client performs decryption to obtain the correct output. Note that ECQCO can be extended to larger-scale quantum circuits as computational resources permit. Circuit-level obfuscation can achieve optimal effectiveness when it is applied to circuits with deterministic outputs. The following subsections provide the implementation details of QCOO and QCSO.

\subsection{Quantum circuit output obfuscation(QCOO)}

According to the concept of QHE \cite{shang2023two}, QCOO enables the trusted client to encrypt and decrypt quantum data using secret keys, while allowing specific quantum computings to be performed directly on the ciphertext without prior decryption. QCOO leverages the homomorphic properties of QOTP encryption to achieve obfuscated computation over the output quantum states. In the decryption phase, we introduce a probabilistic inference technique that allows the recovery of correct measurement outcomes without applying the decryption key. Instead, the client infers the expected result based on the statistical distribution of the obfuscated output. This section describes two crucial technical components of QCOO, namely key generation and update as well as decryption based on reasoning about probability distribution, and then describes the overall scheme design.

The encryption key of QCOO consists of $X$ and $Z$ operators. The $X$, $Z$, $H$, $S$ and $CNOT$ gates are called Clifford group elements, which can maintain the stabilizer state structure \cite{bravyi2016improved}. The Clifford group and $T$ gate form a universal set of quantum gates. For any $n$-qubit Clifford circuit $C$ and any Pauli gate $Q$, there exists another Pauli gate $Q'$ that satisfies $CQ=Q'C$ \cite{gottesman1998heisenberg}. When $Q=X^aZ^b$, $a,b\in\{0,1\}^n$ is used as the key, the key update function of the Clifford gate is represented in Equation \ref{eq1}.
\begin{align}\label{eq1}
    \nonumber f_x(a,b)=(a,b) &, \qquad f_Z(a,b)=(a,b)\\
    f_H(a,b)=(b,a) &, \qquad  f_S(a,b)=(a,a\oplus b)\\
    \nonumber f_{CNOT}(a_1,b_1,a_2,b&_2)=(a_1,b_1\oplus b_2,a_1\oplus a_2,b_2) 
\end{align}

For the $T$ gate and even more generally for any single-qubit gate $U$, it can be represented as $U=e^{i\alpha}R_z(\beta)R_y(\gamma)R_z(\delta)=U(\alpha,\beta,\gamma,\delta)$, through the Z-Y-Z decomposition. The key update function of the $U$ gate is represented in Equation \ref{eq2} \cite{shang2023two}.
\begin{align}\label{eq2}  
    X^aZ^bU(\alpha,\beta,\gamma,\delta)=U(\alpha,(-1)^{a}\beta,(-1)^{a+b}\gamma,(-1)^{a}\delta)X^aZ^b
\end{align}

For any $n$-qubit circuit $C=(g_n,\dots,g_2,g_1)$, where $N$ represents the number of quantum gates in the circuit. the computing party needs to replace $U$ in the circuit according to the key $(a,b)$ and Equation \ref{eq2}, and then $(a,b)$ can be updated. When the circuit acts on the ciphertext quantum state $X^aZ^b|\psi\rangle$, according to the key update function represented in Equation \ref{eq1}, the encryption key $(a_0,b_0)$ can be gradually updated to obtain the decryption key $(a_{final},b_{final})$. The specific update process is represented in Equation \ref{eq3}, and the homomorphic computation result obtained is $X^{a_{final}}Z^{b_{final}}C|\psi\rangle$.
\begin{align}\label{eq3}
   \mathcal{F}_C:\{0,1\}^{2n}\rightarrow\{0,1\}^{2n},f_{g_n}\circ\cdots\circ f_2\circ f_1(a_0,b_0)\rightarrow(a_{final}, b_{final})
\end{align}

In the Clifford+T circuit, since only the $T$ gate in the quantum circuit of the computing party is replaced, using Equation \ref{eq2} to replace the $T$ gate may lead to key leakage. The proof can be found in Appendix A. QCOO calculates the global phase of the quantum circuit to ensure that the computing party does not know which gate is replaced, thus preventing key leakage. Note that the $T$ gate can be written in Equation \ref{eq4}.
\begin{align}\label{eq4}
   T=\left[\begin{array}{cc}1&0 \\0&e^{i\pi/4}\end{array}\right]=e^{i\pi/8}\left[\begin{array}{cc}e^{-i\pi/8}&0 \\0&e^{i\pi/8}\end{array}\right]=e^{i\pi/8}R_z(\pi/4)
\end{align}

Since the global phase $e^{i\pi/8}$ is unmeasurable, gates in the quantum circuit can be replaced with $R_z(\pi/4)$ without affecting the measurement results of the output quantum state. The same conclusion holds for the $T^{\dagger}$ gate. The replacement rule for the $T/T^{\dagger}$ gate is expressed in Equation \ref{eq5}.
\begin{align}\label{eq5}
   T\rightarrow R_z((-1)^a\pi/4),\qquad T^{\dagger}\rightarrow R_z((-1)^a-\pi/4)
\end{align}

Since $a\in\{0,1\}^n$, according to the Equation \ref{eq5}, the set of $T$ gates after replacement is $Set_{T_{gate}}=\{R_z(\pi/4),R_z(-\pi/4)\}$. Due to the randomness of the key, $R_z(\pi/4)/R_z(-\pi/4)$ may be obtained directly from the $T$ gate, or it may be obtained by replacing the $T^{\dagger}$ gate according to the key. Therefore, the computing party cannot infer the original quantum gate from the replaced quantum gate.

In general, quantum circuits that have completed encryption and the replacement of $T/T^{\dagger}$ gates can be correctly decrypted by directly applying the updated $dk$ circuit before measurement. 
In the system model, the compiled quantum circuit must be executed directly on quantum hardware, and the user cannot modify the compiled circuit. To address the constraint, QCOO employs reasoning about probability distribution (RPD) to achieve the decryption functionality.

RPD is based on the reversed application of the \textit{delayed measurement principle} \cite{belavkin1994nondemolition}, as represented in Theorem \ref{t1}. \textit{delayed measurement principle} states that any measurement performed in the middle of a quantum circuit can be postponed to the end, with classical conditional operations replaced by quantum-controlled gates. The same theorem can also be applied in reverse.

\theorem[Reasoning about probability distribution]\label{t1}
{If a quantum circuit $C$ performs measurement only at the final step and yields a probability distribution $P$, then it can be transformed by measuring certain qubits at an intermediate stage of $C$, resulting in a new distribution $P'$. All subsequent quantum operations can then be replaced by classical conditional operations, denoted as $op$. Then there is $P'\overset{op}{\rightarrow}P$
}

The RPD technique relies on three foundations. The first is the determinacy of quantum measurement collapse. The second is the controllability of classical information. The third is the equivalence of measurement outcomes. Note that in real environments, errors like crosstalk and decoherence exist. Excessive use of classical conditional operations to replace quantum noise causes deviation between reconstructed and correct distributions. The deviation becomes pronounced particularly for operations involving superposition, entanglement, or distant measurements.

\begin{algorithm}
\floatname{algorithm}{Algorithm}
\renewcommand{\algorithmicrequire}{\textbf{Input:}}
\renewcommand{\algorithmicensure}{\textbf{Output:}}
\footnotesize
\caption{Quantum circuit output obfuscation algorithm}
\label{alg1}
\begin{algorithmic}[1]
    \REQUIRE The quantum Clifford+T circuit $C$. $C$ consists of $n$ quantum gates, record them in order from left to right as $g_1.g_2,\cdots,g_n$, among which there are $n$ $T/T^{\dagger}$ gates, plaintext (initial) quantum state $|\psi\rangle$. Clifford circuit update rules $f$ according to Equation \ref{eq1}, $T/T^\dagger$ replacement rules $R_{T/T^\dagger}$ according to Equation \ref{eq5};
    \ENSURE Decryption key $dk$ and the quantum circuit $C_{Enc}$ obtained after $C$ encryption;
    \STATE Randomly generate the secret encryption key $ek\Leftarrow(a_0,b_0),a_0,b_0\in\{0,1\}^n$
    \STATE $X^{a_0}Z^{b_0}|\psi\rangle\Leftarrow Enc(ek,|\psi\rangle)$
    \FOR{each gate $g_i\in C$}
        \STATE $C_0\Leftarrow C$
        \IF{$g_i\in\{T/T^\dagger\}$}
            \STATE $C_{i+1} = R_{T/T^\dagger}(C_i,g_i)$;
            \STATE $(a_{i+1},b_{i+1})=(a_i,b_i)$;
        \ELSE
            \STATE $(a_{i+1},b_{i+1})=f_{g_i}(a_i,b_i)$;  
            \ENDIF
        \STATE $C_{Enc}\Leftarrow C_n$
    \ENDFOR
    \STATE $X^{a_{final}}Z^{b_{final}}C|\psi\rangle\Leftarrow Eval^{C_{Enc}}(X^{a_0}Z^{b_0}|\psi\rangle)$
    \STATE $dk\Leftarrow(a_{final},b_{final})$
    \RETURN $dk,C_{Enc}$;
\end{algorithmic}
\end{algorithm}

QCOO adopts RPD because the decryption key operator contains at most $2n$ Pauli operators fixed at the circuit terminus. On one hand, substituted operations are Pauli operators with simple forms. Their finite number ensures low complexity. On the other hand, these operations neighbor final measurements without superposition or entanglement. Noise effects remain limited.

The quantum circuit output obfuscation algorithm is described in Algorithm \ref{alg1}. QCOO algorithm consists of three parts, namely key generation(step 1), encryption(step 2-12) and homomorphic computation(step 13-14).  Decryption is achieved with the help of the RPD. It uses the final key together with the returned measurement results.

\subsection{Quantum circuit structure obfuscation(QCSO)}

By virtue of the concept of QiO \cite{zhang2024quantum}, QCSO enables a trusted client to obfuscate the topological structure and gate-type information of a quantum circuit without altering its computational functionality. The functional equivalence of quantum circuits is achieved by constructing $\Delta$subpath-equivalence. To reduce the computational overhead introduced by structural obfuscation, QCSO analyzes the timing logic of the circuit to locate candidate positions for insertion. Based on the analysis, we design an adaptive decoupling obfuscation algorithm (ADOA). ADOA can take into account both the error suppression of dynamic decoupling and the security protection of the results of the obfuscation circuit.

The following subsections describe the three core techniques of QCSO. These are the construction strategies of $\Delta$subpath-equivalence, ADOA and the probability testing distinguisher. The overall design of the QCSO scheme is then described.

\subsubsection{Construction strategies of $\Delta$subpath-equivalence}
\begin{figure}
    \centering
    \includegraphics[width=0.9\linewidth]{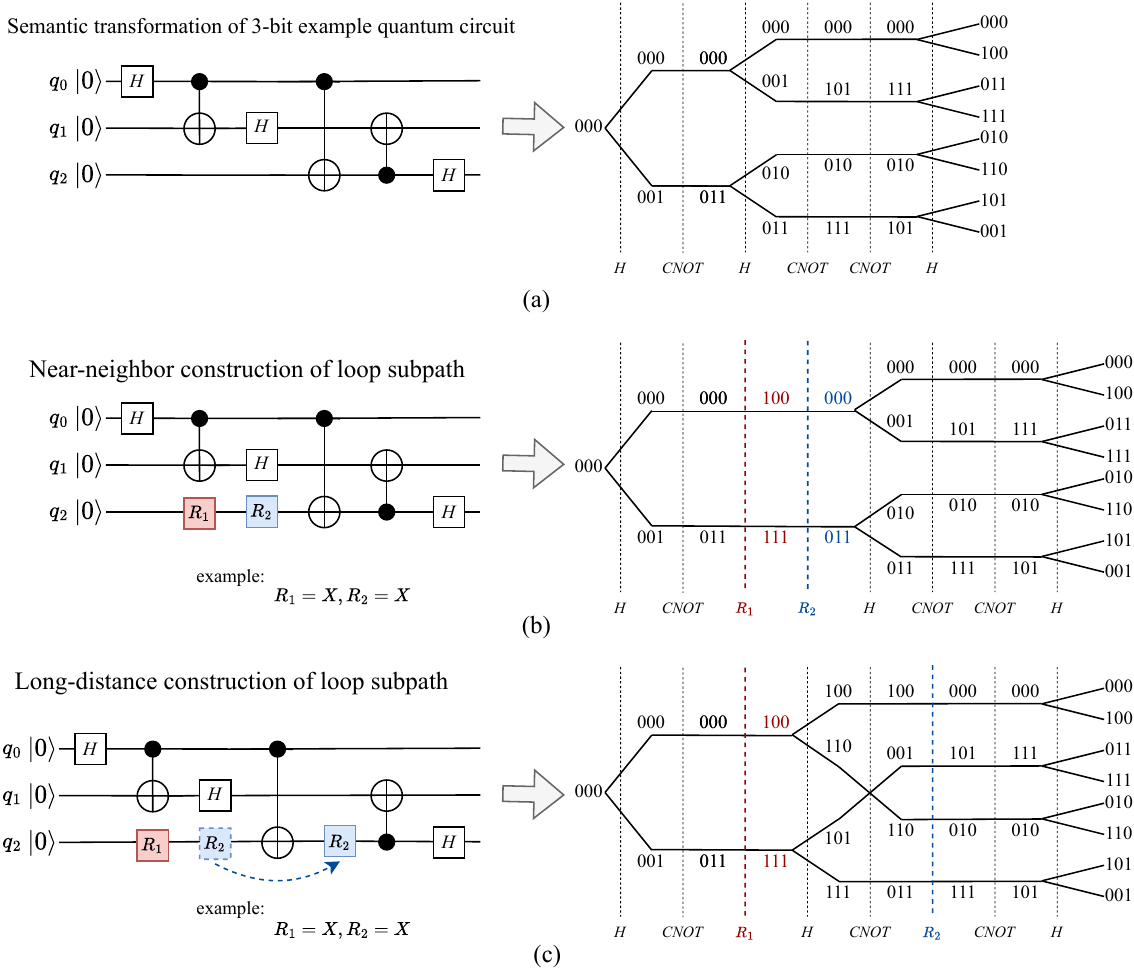}
    \caption{(a) 3-qubit example quantum circuit and its corresponding semantic transformation representation. (b) Quantum circuit for constructing $\Delta$LSP by near-neighbor and its corresponding subpath sum structure. (c) Quantum circuit for constructing $\Delta$LSP by long-distance and its corresponding subpath sum structure}
    \label{subpath}
\end{figure}

In quantum computing, equivalent quantum implementations can perform the same computational task. When given functionally equivalent inputs, the output produced by a copy-protection mechanism becomes computationally indistinguishable. In the recent work \cite{coladangelo2024use}, equivalent quantum implementation is referred to the best copy protection, which also serves as a primary goal of QCSO. QCSO constructs equivalent quantum implementations based on the concept of $\Delta$subpath-equivalence within quantum circuits. The notion originates from the idea of subpath sums in Feynman path integrals \cite{amy2018towards} and is formally defined in Definition \ref{defin1}.

\definition[$\Delta$subpath-equivalence based on subpath sums]\label{defin1}
{Let $C_1$ and $C_2$ be two quantum circuits, and let $SP_1$ and $SP_2$ be their respective subpath sums. The circuits $C_1$ and $C_2$ are said to be $\Delta$subpath-equivalence if there exists a subpath $\Delta SP\subseteq SP$ such that:
\begin{itemize}
    \item The subpath sum operators for the two circuits $C_1$ and $C_2$ are defined as
\begin{equation*}
U_{\Delta SP_{1/2}}=\frac{1}{\sqrt{2^m}}\Sigma_{y\in \mathbf{Z}^m_2}e^{2\pi i\phi_{1/2}(x,y)}|f_{1/2}(x,y)\rangle\langle x|
\end{equation*}
    
    where $x = x_1x_2\cdots x_n$ is the input basis vector (each $x_i$ is a Boolean constant or variable). $y=y_1y_2\cdot y_m$ are the path variables corresponding to intermediate qubits. $\phi_1(x,y)$ and $\phi_2(x,y)$ are the phase polynomials that describe the phase contribution of the subpath sum for $C_1$ and $C_2$. $f_1(x,y)$ and $f_2(x,y)$ are Boolean polynomials describing the output basis states of the circuits.
    \item The operators corresponding to the path sums outside $\Delta SP$ must be identical for both circuits: $U_{\Delta SP_1\notin SP_1} = U_{\Delta SP_2\notin SP_2}$, where $U_{\Delta SP_1\notin SP_1}$ and $U_{\Delta SP_2\notin SP_2}$ are the linear operators defined by the path sums outside the region $\Delta SP$.
    \item The subpath sum operators $U_{\Delta SP_1}$ and $U_{\Delta SP_2}$ must be equivalent: $U_{\Delta SP_1} = U_{\Delta SP_2}$.
\end{itemize}
}

For general quantum circuits, $\phi(x,y)$ may contain high-order terms or non-polynomial forms. If the accumulated phase difference between two paths $U_{\Delta SP_2}$ and $U_{\Delta SP_2}$, satisfies $\Delta\phi(x,y)\equiv0\mathrm{mod}2\pi$, then the two circuits are $\Delta$subpath-equivalence.

A loop subpath $LSP$ refers to a segment of a subpath that forms a closed quantum evolution, where a sequence of unitary operations $U_1,U_2,\dots,U_k$ maps the initial quantum state $|\psi_{init}\rangle$ back to itself, i.e., $U_k,U_{k-1},\cdots,U_1|\psi_{init}\rangle=|\psi_{init}\rangle$. QCSO incorporates $LSP$ into the original subpath segments of a quantum circuit to induce phase cancellation or controlled phase amplification, while ensuring that the resulting circuit and the original circuit remain $\Delta$subpath-equivalence. The modification alters the circuit structure without affecting its computational functionality.

Given a 3-qubit example quantum circuit, as illustrated in Figure \ref{subpath}(a), let the circuit be denoted by $C$, and let $R_1$ and $R_2$ represent a sequence of quantum gates that form a $LSP$, satisfying $R_1R_2|\psi\rangle=|\psi\rangle$. We categorize the construction strategies for $LSP$ into two types. Figure \ref{subpath}(b) illustrates near-neighbor construction, where $R_1$ and $R_2$  are inserted into adjacent positions along the circuit path. The approach temporarily alters the quantum state and then restores it, forming a localized loop. Figure \ref{subpath}(c) illustrates long-distance construction, where $R_1$ and $R_2$ are placed at distant positions in the circuit structure. Despite their separation, they still form a logical loop between the red and blue paths, enabling long-range phase cancellation. Since there may exist multiple valid ways to construct $LSP$, the selection of an appropriate configuration should balance functional equivalence with other considerations such as error suppression, circuit depth, and computational overhead.

\subsubsection{Adaptive decoupling obfuscation algorithm}
According to the concept of dynamic decoupling \cite{das2021adapt}, QCSO constructs $LSP$ through the Adaptive Decoupling Obfuscation Algorithm (ADOA). The use of two-qubit gates will incur considerable overhead and may cause crosstalk errors. Given the durations of a set of universal quantum gates, ADOA obtains the idle positions of the quantum circuit $C$ under analog operation through discrete-to-analog frame conversion based on the given quantum circuit $C$. 

\begin{algorithm}
\renewcommand{\algorithmicrequire}{\textbf{Input:}}
\renewcommand{\algorithmicensure}{\textbf{Output:}}
\footnotesize
\caption{Adaptive decoupling obfuscation algorithm}
\label{alg2}
\begin{algorithmic}[1]
    \REQUIRE The quantum circuit $C$, the durations of a set of universal quantum gates $S_{duration}$, an empty set of circuit idle positions $Free$, an empty instruction list $L_{empty}$. In $C$, the set of single-qubit gates for the near-neighbor before and after the idle position is $\{g_{context}\}$, obfuscation decoupling parameters $\lambda$;
    \ENSURE The quantum circuit after QCSO$\Rightarrow C_{QCSO}$;
    \STATE $DAG_C\Leftarrow getDAGgraph(C)$;
    \STATE Populate $L_{empty}$ with operations from $DAG_C$, obtain the discrete frames of $C\Rightarrow Df_C$; 
    \STATE Convert discrete frames into analog frames, $Af_C\Leftarrow convert(Df_C,S_{duration})$;
    \FOR{each analog frame $af\in Af_C$}
        \FOR{each qubit $q_i\in C$}
            \STATE Calculate the idle duration and position, denoted as $t_i,p_i$, respectively;
            \IF{$t_i > 0$}
                \STATE Add $p_i$ to $Free_i$ and merge adjacent $p_i$ in time;
            \ENDIF
        \ENDFOR
    \ENDFOR
    \FOR{each qubit $q_i\in C$}
        \FOR{each idle position $p_j\in Free_i$}
            \IF{$p_j > XY-8$}
                \STATE Insert $XY-8$ sequence at $p_j$, obtain $C_{ij}$, $C_{QCSO}\Leftarrow C_{ij}$;
            \ELSIF{$p_j > XY-4$}
                \STATE Insert $XY-4$ sequence at $p_j$, obtain $C_{ij}$, $C_{QCSO}\Leftarrow C_{ij}$;
            \ELSIF{$p_j > XX$}    
                \STATE Insert $XY-4$ sequence at $p_j$, obtain $C_{ij}$, $C_{QCSO}\Leftarrow C_{ij}$;
            \ELSIF{$p_j > Z$ and $p_j < XX$ and $\{g_{context}\}\ne\emptyset$ and $\lambda=True$}
                \STATE Insert $Z$ sequence at $p_j$, combine $Z$ and $g_{context}$ into a new $U3$ gate, obtain $C_{ij}$, $C_{QCSO}\Leftarrow C_{ij}$;
            \ENDIF
        \ENDFOR
    \ENDFOR
    \RETURN $C_{QCSO}$

\end{algorithmic}
\end{algorithm}

To reduce the impact of additional gates on compilation and execution performance, ADOA applies a periodic series of inversion pulses ($XX,XY-4/8$) to the quantum bits. The gates $R_1$ and $R_2$ that form the $LSP$ are inserted into idle positions within the quantum circuit, which serves to suppress idle-time decoherence. The approach corresponds to the adjacent construction of $LSP$ mentioned above. If the idle position is insufficient to insert the minimum pulse sequence, then ADOA will check whether there are adjacent single-qubit gates before and after this position. If such gates exist, a $ZZ$ pulse is inserted, and one of the $Z$ gates is combined with an adjacent qubit gate to form a new single-qubit gate. If no adjacent single-qubit gates are present, the circuit remains unchanged. The approach is an alternative to the long-distance construction of $LSP$. When $LSP$ is constructed over large-scale circuits at a long distance, it will generate a huge amount of computation. An "approximate" long-distance construction can be achieved by inserting $Z$ gates at multiple small idle positions. The reason for inserting the $ZZ$ sequence instead of the XX sequence is that in the merging scenario, the $ZZ$ sequence pair has better performance in suppressing dephasing noise and crosstalk residues. It is easier to maintain the logic after merging.

Although the insertion of any pulse sequence contributes positively to the obfuscation of the quantum circuit structure, the merging of $ZZ$ sequences weakens the suppression of decoherence noise. Therefore, ADOA sets the obfuscation decoupling parameters to achieve a trade-off between noise suppression and circuit structure protection. The detailed procedure of ADOA is described in Algorithm \ref{alg2}.

\subsubsection{Probability testing distinguisher}

Although QCSO inserts pulse sequences that are mathematically equivalent, we still need to verify whether the scheme preserves quantum indistinguishability in functionality. It is necessary for establishing both correctness and the achievable security level. One natural approach is to test all possible inputs. As the input size increases, the number of possible inputs increases exponentially. It leads to high computational cost and potential loss of security guarantees.

QCSO uses a method called probabilistic testing distinguisher (PTD). PTD is based on the idea of polynomial identity testing under semantic optimization, which can reduce the indistinguishability verification problem to the equivalence of $SP$. PTD randomly samples the path variables of the quantum circuit and checks whether it satisfies $\Delta$subpath-equivalence. If they are not equal, the test finds a counterexample. If they are equal, the two quantum circuits are considered functionally equivalent with high probability. 

\subsection{ECQCO scheme}

\begin{figure}
    \centering
    \includegraphics[width=1\linewidth]{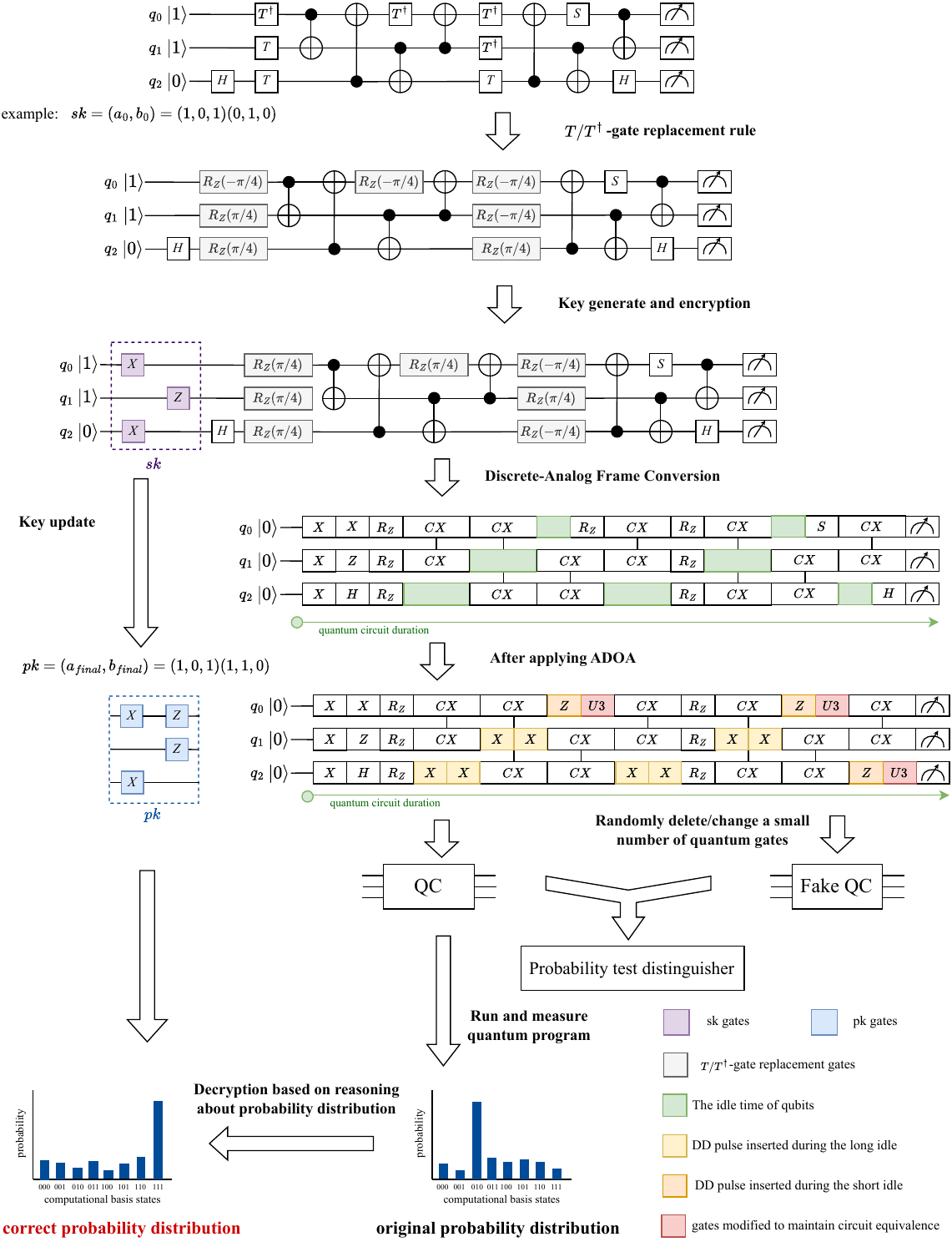}
    \caption{The process of applying ECQCO to the Toffoli gate decomposition circuit}
    \label{ecqco_example}
\end{figure}

To better illustrate how ECQCO encrypts the quantum circuit at the user end, we take the Toffoli gate decomposition circuit as an example to introduce the ECQCO scheme, as illustrated in Figure \ref{ecqco_example}. Assume that the duration of all single-qubit gates is the same and the $CX$ gate is exactly twice time that of the single-qubit gate, although it is not necessarily the case in reality.

In Figure \ref{ecqco_example}, ECQCO will apply the QCOO algorithm (Algorithm \ref{alg1}) to the circuit $C$ firstly. Assume that the randomly generated key is $sk=(a_0,b_0)=(1,0,1)(0,1,0)$ (the purple circuit). The $T/T^\dagger$ gates in $C$ are replaced by $R_Z(\pi/4)/R_Z(-\pi/4)$ gates (the grey gates above). Update the key according to the quantum gate information in $C$ to obtain the decryption key $pk=(a_{final},b_{final})=(1,0,1)(1,1,0)$ (blue circuit). At the same time, the change of the key will also modify the replaced $R_Z$ gate (the gray gate below). The quantum circuit that completes the update and replacement, together with $sk$, constitutes the encryption circuit $C_{Enc}$.

Subsequently, $C_{Enc}$ goes through a discrete-analog frame conversion to obtain all the idle positions (green squares) in the circuit. According to the ADOA (Algorithm \ref{alg2}), a pulse sequence is inserted into the idle positions (the example assumes $\lambda$ is true). The $XX$ sequences (yellow gate) are inserted into long idles, and the $ZZ$ sequences (orange gate) are inserted into short idles and merged with the near-neighbor single-qubit gates to obtain the corresponding $U3$ gate (red gate), resulting in the scrambled circuit $QC$. A copy of $QC$ has a small number of quantum gates randomly deleted/changed (5\% - 15\%) to obtain $Fake\_QC$. $QC$ and $Fake\_QC$ are subjected to equivalence verification through PTD. After the verification is correct, $QC$ is run and measured to obtain the original probability distribution. Finally, with the help of the RPT, the original probability distribution is restored to the correct probability distribution according to $pk$, thus completing the encrypted-state quantum compilation.

\subsection{Correctness and security analyses}\label{sec4}
The correctness of the ECQCO scheme consists of the combined correctness guarantees of QCOO and QCSO. QCOO is $\mathcal{F}$-homomorphic, as represented in Theorem \ref{t2}. The theoretical correctness of QCSO comes from the Schwartz–Zippel lemma \cite{motwani1996randomized}, and it is verified experimentally through positive and negative testing in Section \ref{sec4}.

\subsubsection{Correctness analysis}

\theorem[The correctness of QCOO]{QCOO is $\mathcal{F}$-homomorphic.}\label{t2}

\begin{proof}
    The definition of $\mathcal{F}$-homomorphic is given in Definition \ref{defin_f}. Any quantum circuit can be constructed by Clifford+$T$ gates. Without loss of generality, we consider an $n$-qubit quantum circuit $C\in \mathcal{F}$ that contains at least one $T/T^\dagger$ gate. Suppose the first $T$ gate $g_{i,j}$ is the j-th quantum gate, acting on the i-th qubit, i.e, $g_{i,j}=T$. $C$ can be expressed as $C=\Omega_2T/T^\dagger \Omega_1$, where $\Omega_1$ contains only Clifford gates, and $\Omega_2$ consists of Clifford+$T/T^\dagger$.

    The user encrypts the plaintext state $|\psi\rangle=|\alpha\rangle\otimes|\omega\rangle\otimes|\beta\rangle$ with QOTP, which produces a ciphertext state $X^{a_0}Z^{b_0}|\psi\rangle=X^{\otimes^n_{k=1}a_0(k)}Z^{\otimes^n_{k=1}b_0(k)}(|\alpha\rangle\otimes|\omega\rangle\otimes|\beta\rangle)$. During the key update process, after updating $\Omega_1$, the key is $(a_{j-1},b_{j-1})$. When updating the first $T$ gate of $C$, firstly replace the $T$ gate with $R_Z\big((-1)^{a_{j-1}(i)}\pi/4\big)$, and then update the key $(a_{j},b_{j}) = (a_{j-1},b_{j-1})$. The replaced quantum circuit $C_{mid}$ is $\Omega_2(I_{i-1}\otimes R_Z\big((-1)^{a_{j-1}(i)}\pi/4\big)I_{n-i})\Omega_1$. At the time, when the quantum circuit $C$ acts on the encrypted quantum state, Equation \ref{eq_cmid} holds. 
    \begin{align}\label{eq_cmid}
        C_{mid}X^{a_0}Z^{b_0}|\psi\rangle&=\Omega_2\Big(I_{i-1}\otimes R_Z\big((-1)^{a_{j-1}(i)}\pi/4\big)I_{n-i}\Big)\Omega_1X^{a_{0}}Z^{b_{0}}|\psi\rangle
    \end{align}

    According to the Clifford gate key update function, the key updated after the operation $\Omega_1$ is $(a_{j-1},b_{j-1}) = \Omega(a_{0},b_{0})$. Therefore, Equation \ref{eq_omega1} holds after the operation $\Omega_1$.
    \begin{align}\label{eq_omega1}
        C_{mid}X^{a_0}Z^{b_0}|\psi\rangle=&\Omega_2\Big(I_{i-1}\otimes R_Z\big((-1)^{a_{j-1}(i)}\pi/4\big)I_{n-i}\Big)X^{a_{j-1}}Z^{b_{j-1}}\Omega_1|\psi\rangle\\\nonumber 
        =&\Omega_2\Big(I_{i-1}\otimes R_Z\big((-1)^{a_{j-1}(i)}\pi/4\big)I_{n-i}\Big)\\\nonumber &
        \Big(X^{\otimes^{i-1}_{k=1}a_{j-1}(k)}Z^{\otimes^{j-1}_{k=1}b_{j-1}(k)}\otimes X^{a_{j-1}(i)}Z^{b_{j-1}(i)}\otimes X^{\otimes^{n}_{k=i+1}a_{j-1}(k)}Z^{\otimes^{n}_{k=i+1}b_{j-1}(k)}\Big)
        \Omega_1|\psi\rangle
    \end{align}

    According to the properties of $R_Z$, Equation \ref{eq_rz} holds.
    \begin{align}\label{eq_rz}
        R_Z\big((-1)^{a_{j-1}(i)}\pi/4\big)X^{a_{j-1}(i)}Z^{b_{j-1}(i)}=X^{a_{j-1}(i)}Z^{b_{j-1}(i)}R_Z(\pi/4)
    \end{align}

    According to the absorption law of the tensor product $(A\otimes B)(C\otimes D)=AC\otimes BD$ and Equation \ref{eq_rz}, Equation \ref{eq_omega2} holds.
    \begin{align}\label{eq_omega2}
    &\Big(I_{i-1}\otimes R_Z\big((-1)^{a_{j-1}(i)}\pi/4\big)\otimes I_{n-i}\Big)\\\nonumber&
    \Big(X^{\otimes^{i-1}_{k=1}a_{j-1}(k)}Z^{\otimes^{j-1}_{k=1}b_{j-1}(k)}\otimes X^{a_{j-1}(i)}Z^{b_{j-1}(i)}\otimes X^{\otimes^{n}_{k=i+1}a_{j-1}(k)}Z^{\otimes^{n}_{k=i+1}b_{j-1}(k)}\Big)\\\nonumber 
    =& X^{\otimes^{i-1}_{k=1}a_{j-1}(k)}Z^{\otimes^{j-1}_{k=1}b_{j-1}(k)}\otimes 
    R_Z\big((-1)^{a_{j-1}(i)}\pi/4\big)X^{a_{j-1}(i)}Z^{a_{j-1}(i)}\otimes
     X^{\otimes^{n}_{k=i+1}a_{j-1}(k)}Z^{\otimes^{n}_{k=i+1}b_{j-1}(k)} \\\nonumber 
    =& X^{\otimes^{i-1}_{k=1}a_{j-1}(k)}Z^{\otimes^{j-1}_{k=1}b_{j-1}(k)}\otimes 
    X^{a_{j-1}(i)}Z^{a_{j-1}(i)}R_Z(\pi/4)\otimes
     X^{\otimes^{n}_{k=i+1}a_{j-1}(k)}Z^{\otimes^{n}_{k=i+1}b_{j-1}(k)} \\\nonumber 
    =& \big(X^{\otimes^{i-1}_{k=1}a_{j-1}(k)}Z^{\otimes^{j-1}_{k=1}b_{j-1}(k)}\otimes X^{a_{j-1}(i)}Z^{b_{j-1}(i)}\otimes X^{\otimes^{n}_{k=i+1}a_{j-1}(k)}Z^{\otimes^{n}_{k=i+1}b_{j-1}(k)}\big)
    \big(I_{i-1}\otimes R_Z(\pi/4)\otimes I_{n-i}\big)\\\nonumber 
    =& X^{a_{j-1}}Z^{a_{j-1}}\big(I_{i-1}\otimes R_Z(\pi/4)\otimes I_{n-i}\big)
    \end{align} 
    
    The key remains unchanged after the action of the $T$ gate, satisfying $(a_j,b_j)=(a_{j-1},b_{j-1})$.Therefore, Equation \ref{eq_10} holds.
    \begin{align}\label{eq_10}
        C_{mid}X^{a_0}Z^{b_0}|\psi\rangle= \Omega_2X^{a_{j}}Z^{a_{j}}\big(I_{i-1}\otimes R_Z(\pi/4)\otimes I_{n-i}\big)\Omega_1
    \end{align}

    The computing party can complete the quantum homomorphic encryption of $\Omega_1$ and the first $T$ gate, according to Equation \ref{eq_10}. The same applies to the $T^\dagger$ gate. Similarly, the quantum homomorphic encryption of $\Omega_2$ can be completed according to the above process. After the encryption is completed, $CX^{a_0}Z^{b_0}|\psi\rangle=X^{a_{final}}Z^{b_{final}}C|\psi\rangle$  holds. By applying $dk=(a_{final},b_{final})$ to construct the decryption operator $Z^{b_{final}}X^{a_{final}}$, the correct plaintext result $C|\psi\rangle$ is obtained. Therefore, QCOO is $\mathcal{F}$-homomorphic.

\end{proof}

The correctness of QCSO relies on verifying the obfuscated quantum circuit using the Probabilistic Testing Distinguisher (PTD). The verification is based on the extended semantic transformation of quantum implementations \cite{xu2023synthesizing} and the Schwartz–Zippel lemma \cite{motwani1996randomized}. The proof can be found in Appendix B. In practice, we adopt the widely used \textit{positive-negative testing} from classical integrated circuit design. The positive test checks whether the circuit remains functionally equivalent after obfuscation. The negative test introduces changes to the obfuscated circuit by randomly adding or removing 5\% to 15\% of selected quantum paths. It then re-evaluates the functional equivalence. A single counterexample is sufficient to determine inequality, making the test verifiable in polynomial time. By comparing the time costs of the positive and negative tests in the experiments (Section \ref{sec4.2}), we reduce the overall verification complexity from exponential to polynomial scale.

\subsubsection{Security analysis}
The security of the ECQCO scheme consists of the combined security guarantees of QCOO and QCSO. QCOO achieves information-theoretic security, as represented in Theorem \ref{t3}. QCSO is quantum indistinguishable secure, under the quantum random oracle, as represented in Theorem \ref{t4}.
\theorem[The security of QCOO]{QCOO is information-theoretically secure.}\label{t3}

\begin{proof}
    The user encrypts the plaintext state $|\psi\rangle$ using QOTP, which produces a ciphertext state $X^{a_0}Z^{b_0}|\psi\rangle$ that is a maximally mixed state. As a result, the computing server cannot obtain any information about the $|\psi\rangle$ or $(a_0,b_0)$. The replacement of quantum gates within the circuit does not reveal any information about the key. The computing server cannot infer the intermediate key values and cannot derive the $(a_{final},b_{final})$. The scheme completely hides both the input and the output. In addition, the security of QOTP and gate replacement does not rely on any computational assumptions. Thus, QCOO achieves information-theoretic security.
\end{proof}

\theorem[The security of QCSO]{QCSO is quantum indistinguishable secure, under the quantum random oracle.}\label{t4}

\begin{proof}
    We assume that there exists two quantum implementations $(\rho_0,C_0),(\rho_1,C_1)$ of a classical function $f$, defined in Definition \ref{df_qi}.
    \definition[Quantum implementation of classic function]\label{df_qi}
    {
    Let $n,m\in\mathbf{N}$, classic function $f:\{0,1\}^n\rightarrow\{0,1\}^n,\epsilon\in[0,1]$.The $(1-\epsilon)$-quantum implementation of $f$ is a pair $(\rho,C)$, $\rho$ is the quantum state of the system and $C$ is the quantum circuit that satisfies Equation \ref{eq_qi}.
    \begin{align}\label{eq_qi}
        \forall x\in\{0,1\}^n,\quad\mathrm{Pr}[C(\rho,x)=f(x)]\geq 1-\epsilon
    \end{align}
    }
    If $(\rho_0,C_0)$ and $(\rho_1,C_1)$ satisfy Equation \ref{eq_eq}, then we say that $(\rho_0,C_0)$ and $(\rho_1,C_1)$ are two equivalent quantum implementations of $f$. 
    \begin{align}\label{eq_eq}
        |\mathrm{Pr[D(\rho_0,C_0)=1]}-\mathrm{Pr[D(\rho_1,C_1)=1]}|\leq negl(\lambda)
    \end{align}
    QCSO inserts the identity gate into $(\rho_0, C_0)$ to obtain $(\rho_1, C_1)$, where each inserted sequence forms $LSP$ and satisfies the $\Delta$subpath equivalence. Apart from the inserted sequences, the circuits remain exactly the same, and the subpath sum is preserved. So the $(\rho_0,C_0)$ and $(\rho_1,C_1)$ after the action of QCSO are two equivalent quantum implementations. According to Definition \ref{def_qio}, a QiO scheme can be constructed based on quantum circuit equivalence if equivalent quantum implementations are satisfied. In this way, the security of QCSO can be attributed to QiO \cite{coladangelo2024use}, and the universal security of QiO is the quantum indistinguishability under the quantum random oracle. Note that the derivation of the security proof for QiO is too long and not the focus of our article. More technical details can be found in \cite{zhang2024quantum}.
\end{proof}

\section{Experiments}\label{sec4}

\subsection{Experiment setup}\label{sec4.1}

The scheme is implemented by Python 3.11, leveraging Qpanda3 \cite{zou2025qpanda3} for simulating quantum compilation and operation. The experiments were conducted on a Windows 11 system equipped with an Intel i7-12700H CPU and an NVIDIA GeForce RTX 3060 GPU. The benchmark circuits were selected from the standard library \cite{burgholzer2020advanced,peham2022equivalence} constructed with “high-level” descriptions in RevLib\cite{wille2008revlib}, as well as reconstructed implementations of representative quantum algorithms. These benchmarks include reversible arithmetic circuits and rigorous implementations of quantum algorithms. They have been widely adopted in prior work \cite{burgholzer2020advanced,peham2022equivalence,zhang2024quantum} on quantum circuit compilation and equivalence verification. These benchmarks allow us to comprehensively evaluate the scalability of our system across a range of circuit complexities.

To ensure the reality of our experiments, we used the \textit{core.NoiseModel} module in Qpanda \cite{zou2025qpanda3} to construct a noise-aware quantum simulation environment, which integrates various noise models derived from the \textit{Wukong} 72-qubit superconducting quantum computer developed by OriginQ. We set up a simulation environment with readout noise, decoherence noise, and CZ gate errors, while neglecting single-qubit gate noise. We use the quantum gate duration calculation of the Quafu cloud platform \cite{jin2024quafu}. To better demonstrate the effectiveness of the scheme, we compare ECQCO with several representative quantum circuit obfuscation methods, including inverse gates \cite{das2023randomized}, composite gated \cite{bartake2025obfusqate}, and delayed gates \cite{bartake2025obfusqate}. The comparison covers multiple aspects of circuit transformation and obfuscation capability. 

\subsection{Correctness}\label{sec4.2}

The correctness of ECQCO relies on validating both QCOO and QCSO. Since the verification complexity of QCOO is polynomial, we can efficiently test the consistency between the decrypted output and the original plaintext through experiments. In contrast, QCSO requires exponential resources for full verification, so PTD is applied to assess functional equivalence after obfuscation.

\begin{table}[!t]
\footnotesize
\caption{Verification Results after ECQCO}
\label{tab1}
\tabcolsep 14pt 
\begin{tabular*}{\textwidth}{ccccccc}
\toprule
  \multirow{2}{*}{Benchmarks} & \multirow{2}{*}{qubits} & \multirow{2}{*}{path variables} & \multirow{2}{*}{Clifford gates} & \multirow{2}{*}{$T$-gates} & \multicolumn{2}{c}{Time(s)} \\
&  &  & & & Positive & Negative\\\hline
$\mathrm{Toffoli_3}$ & 5 &12 & 52 &36  & 0.002 &0.001\\
$\mathrm{Toffoli_{10}}$ & 19 & 68 & 297& 190 &0.034& 0.051\\
$\mathrm{VBE\_Adder_3}$ & 10  & 20 & 167 & 94 & 0.021 &0.017\\
$\mathrm{Toff\_Barenco_3}$ & 5 & 12  & 66 & 44 & 0.002 &0.002\\
$\mathrm{Toff\_Barenco_{10}}$  &19 &68 & 493 & 324 &0.093 &0.078\\
$\mathrm{RC\_Adder_6}$  &14 &44 &322 &124 & 0.097 & 0.059\\
$\mathrm{Adder_8}$  & 24 & 160  & 1419 & 614 & 3.732 & 4.186 \\
$\mathrm{Grover_5}$ &9& 200 &1515 &490 & 1.035 & 0.934\\
$\mathrm{Mod\_Adder_{1024}}$ &28 &660 &4363 &3006 & 73.59 & 63.128 \\
$\mathrm{QCLA\_Mod_7}$ &26 &164 &1641 &650 & 47.523 & 51.274\\
$\mathrm{QFT_4}$ &5& 84& 218 &136 & 0.05 & 0.056 \\
$\mathrm{Hamming_{15}}$ &20 &716 & 5332 & 3462 & 86.868 & 90.423\\
$\mathrm{HWB_6}$ & 7 & 52 & 369 & 180 & 0.205 & 0.241\\
$\mathrm{CSUM\_MUX_9}$ & 30 & 56 & 638 & 280 & 0.474 & 0.581\\
$\mathrm{GF(2^4)\_Mult}$ &12 &28& 263 &180& 0.01 & 0.013\\
$\mathrm{GF(2^8)\_Mult}$ &24 &60 &975 &712 &0.089 &0.11\\
$\mathrm{GF(2^{16})\_Mult}$ &48 &124& 3694 &2832& 1.24 & 0.969\\
$\mathrm{GF(2^{32})\_Mult}$ &96 &252& 14259 & 11296& 14.21 & 15.725\\
$\mathrm{GF(2^{64})\_Mult}$ &192 &508 &55408 &45120& 129.135 & 137.823\\
$\mathrm{GF(2^{128})\_Mult}$ &384& 1020 & 231318 & 180352 & 2308.857 & 2195.42\\
\bottomrule
\end{tabular*}
\end{table}

Table \ref{tab1} lists the verification results of quantum circuits obfuscated by ECQCO. The columns labeled Clifford gates and $T$-gate indicate the number of Clifford and $T$ gates, respectively. The Positive and Negative columns report the time required to confirm functional equivalence and non-equivalence, respectively. As listed in Table \ref{tab1}, all obfuscated benchmark circuits passed the functional equivalence test, resulting in a 100\% success rate. The largest circuit contains 384 qubits, 1020 path variables, and more than 410000 gates. It completed verification in approximately 38 minutes. All other benchmarks completed verification within 2 minutes, and 55\% of them finished in under 1 second. The time difference between reverse verification and forward verification is within approximately 6\%. It indicates that ECQCO successfully reduces the equivalence verification to the polynomial level $O(n)$, thus verifying the correctness of QCSO.

\subsection{Obfuscation effect}

Total variation distance (TVD) is a standard metric in probability theory for quantifying the difference between two probability distributions. It has been widely used in quantum circuit obfuscation research. TVD is computed as the sum of the absolute differences between the output counts of the obfuscated and original circuits, normalized by the total number of shots. TVD value closer to 1 indicates a greater ability of QCOO to alter the output distribution of the circuit, but it also implies a higher correlation between the obfuscated and original outputs. When TVD is kept at a relatively high level, the obfuscation effect is preserved while the correlation is reduced to some extent. TVD is defined by Equation \ref{eq_tvd}, where $N$ represents the total number of shots in this run, $n$ is the number of bits in the output, $y_{ECQCO_i}$ and $y_{original_i}$ represent the total number of measurement outcomes of value $i$ in the ECQCO and original quantum circuits, respectively.
\begin{align}\label{eq_tvd}
    \mathrm{TVD}=\frac{\sum^{2^n-1}_{i=0}|y_{ECQCO_i}-y_{origin_i}|}{2N}
\end{align}

The normalized graph edit distance (normGED) is a classical metric used to measure structural differences between two graphs. It computes the minimum total cost required to transform one graph into another by applying a set of defined edit operations, and normalizes the cost by the maximum possible value. NormGED value closer to 1 indicates a more substantial structural transformation, which suggests that the quantum circuit structure obfuscation is stronger. Given two graphs $G_1=(V_1,E_1),G_2=(V_2,E_2)$, the graph edit distance is defined as the sum of the minimum edit costs required to transform $G_1$ into $G_2$, including add/delete/replace nodes and edges. $GED(G_1,G_2)$ is denoted as the minimum total cost and the maximum possible graph edit distance is represented as $\mathrm{maxGED(G_1,G_2)}$. normGED is defined by Equation \ref{eq_nGED}
\begin{align}\label{eq_nGED}
    \mathrm{normGED} = \frac{\mathrm{GED}(G_1,G_2)}{\mathrm{maxGED}(G_1,G_2)}=\frac{\mathrm{GED}(G_1,G_2)}{\mathrm{max}(|V_1|,|V_2|)+\mathrm{max}(|E_1|,|E_2|)}
\end{align}

We have selected four quantum algorithms to measure the overheads of different algorithms and schemes, including the Bernstein-Vazirani algorithm, the Grover algorithm, the Quantum Approximate Optimization Algorithm (QAOA), and Shor’s algorithm. TVD and normGED are used to evaluate how ECQCO impacts the output distribution and structural topology of the circuits, respectively. The encryption key of ECQCO is randomly selected, resulting in different quantum circuits for each obfuscation. Therefore, in the experimental data, the ECQCO-related indicators represent the average values obtained after 10 measurements.

\begin{figure}
    \centering
    \includegraphics[width=1\linewidth]{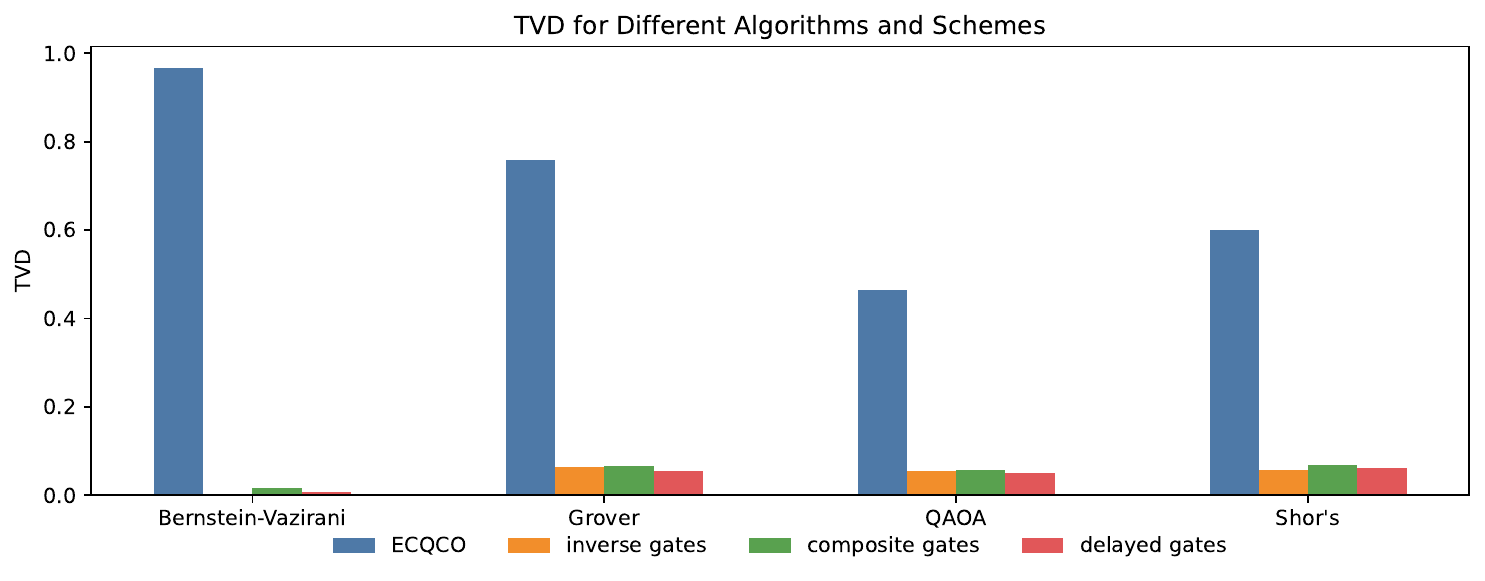}
    \caption{Total variation distance from circuit-based obfuscation}
    \label{tvd}
\end{figure}

\begin{figure}
    \centering
    \includegraphics[width=1\linewidth]{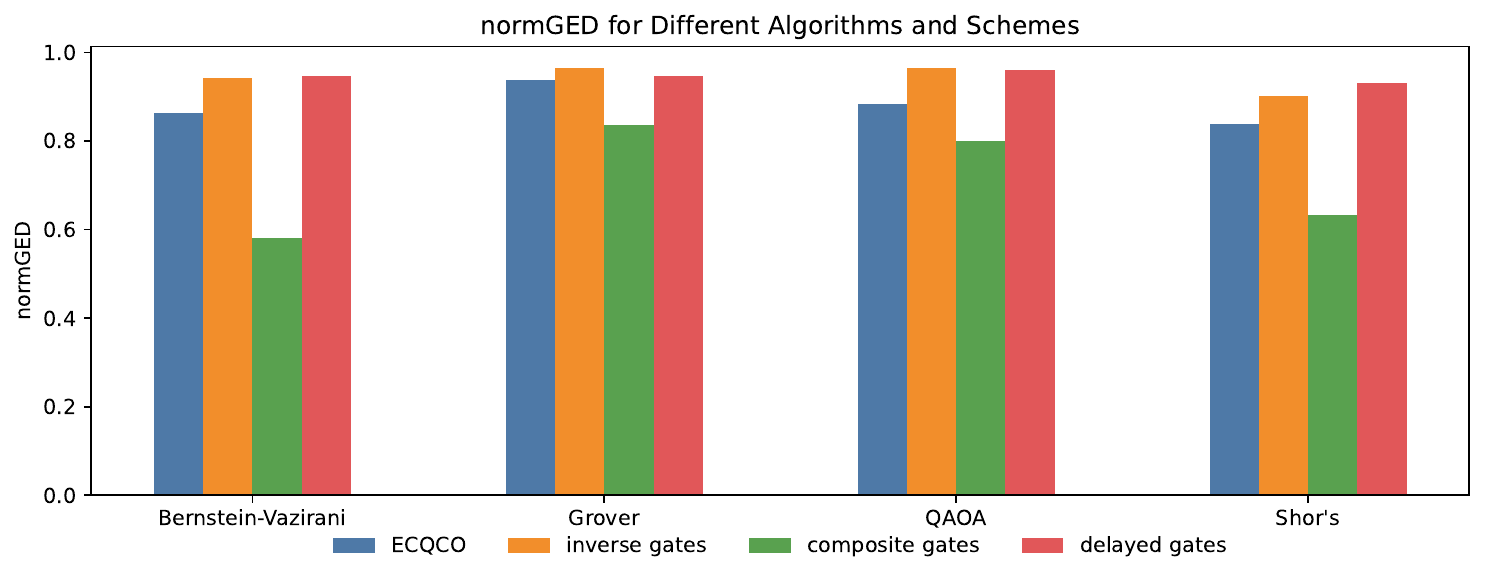}
    \caption{The normalized graph edit distance from circuit-based obfuscation}
    \label{nGED}
\end{figure}

Figures \ref{tvd} and \ref{nGED} show the TVD and normGED results under different schemes for some common quantum algorithms, respectively. After applying ECQCO, the average TVD can reach 0.7, while normGED can reach a relatively high level of 0.88. It indicates that ECQCO effectively obfuscates both output and structure across common quantum programs. TVD value below 1 indicates ECQCO can produce neutral outputs, making it harder for adversaries to infer the circuit functionality. In contrast, other related schemes also maintain high normGED values but show only limited improvements in TVD, as they primarily focus on structural changes with limited protection for output behavior.

\subsection{Overhead and Fidelity}
Security-aware quantum compilation requires a balance between protection and efficiency. Excessive insertion of quantum gates or ancillary qubits contradicts the fundamental goals of quantum compilation. Table 1 presents the depth, analog-frame-based runtime, and fidelity of representative quantum algorithms under different circuit protection schemes. Due to the use of encrypted quantum circuits introduced by the output obfuscation mechanism, ECQCO slightly increases the circuit depth, and the total runtime grows by an average of 3\% compared to the original circuits. Since the structure encryption in ECQCO adopts fixed-depth circuits, the overhead in runtime becomes even less significant as the circuit scales. As shown in Table 1, the fidelity variation after ECQCO transformation remains within 1\% across most algorithms, and even improves by up to 5\% for the Bernstein–Vazirani algorithm. The improvement is attributed to the dynamic decoupling mechanism embedded in ECQCO, which suppresses idle-time decoherence errors.

While the composite gates scheme also introduces modest increases in depth and runtime, it requires doubling the number of auxiliary qubits for gate merging, which enlarges the quantum volume of the compiled circuit and moderately reduces fidelity. In contrast, the inverse gates and delayed gates schemes introduce a large number of additional quantum gates, which significantly increase both the circuit depth and its duration. As a result, these schemes suffer from intensified decoherence noise and lead to lower overall fidelity.

\begin{table}[!t]
\footnotesize
\caption{Comparison among different quantum circuit protection schemes}
\label{tab2}
\tabcolsep 9pt 
\renewcommand{\arraystretch}{1.3}
\begin{tabularx}{\linewidth}{
        >{\centering\arraybackslash}p{2.5cm}  
        *{6}{>{\centering\arraybackslash}X}   
    }
\toprule
\multicolumn{2}{c}{\diagbox[dir=SE]{Algorithms}{Schemes}} &original & inverse gates\cite{das2023randomized} & composite gates\cite{bartake2025obfusqate} & delayed gates\cite{bartake2025obfusqate} &  ECQCO\\\hline
 
\multirow{3}{*}{Bernstein-Vazirani} & Depth &	14	& 182 & 12 & 183 & 16\\
	& Duration(ms)	&1494 & 26159 & 1591 & 22490 & 1662\\
        &	Fidelity	& 0.9029 & 0.136 & 0.9117 & 0.2076 & 0.9597\\\hline
\multirow{3}{*}{Grover} & Depth &	20	& 187 & 20 & 182 & 22\\
	&Duration(ms)	&	2668	& 26159 & 2164 & 22406 & 2736\\
        &	Fidelity	& 0.9930 & 0.3809 & 0.8526 & 0.3835 & 0.9863\\\hline
\multirow{3}{*}{QAOA} & Depth &15 & 171 & 16 & 154 & 16\\
	&Duration(ms)	& 2504 & 25001 & 2276 & 19953 & 2588\\
        &	Fidelity	&0.9952 & 0.8357 & 0.9835 & 0.755 & 0.9753 \\\hline
\multirow{3}{*}{Shor's} & Depth &	14 & 185 & 16 & 189 & 15\\
	&Duration(ms)	& 2284 & 27102 & 2166 & 23887 & 2302\\
        &	Fidelity	&0.9884 & 0.8669 & 0.9618 & 0.8864 & 0.9861\\\hline

\bottomrule
\end{tabularx}
\end{table}


\section{Conclusion}\label{sec5}
In this work, we proposed a quantum encrypted-state compilation scheme based on quantum circuit obfuscation. The scheme leveraged efficiently instantiated quantum indistinguishability obfuscation and quantum homomorphic encryption to protect both the output and structural information of quantum circuits. It achieved a strong balance between security and efficiency by building on quantum cryptographic primitives. It introduced only slight increases in circuit complexity, with average fidelity variation remaining within 1\%. Experimental results demonstrated that our method was well-suited for quantum cloud compilation scenarios in the NISQ era, especially where quantum program privacy was required.

Furthermore, the effectiveness of our approach for large-scale quantum programs and hybrid quantum-classical algorithms with frequent classical interaction (such as multi-layer QAOA) remains to be further explored. While the theoretical security guarantees remained valid, the practical realization of encrypted-state compilation requires additional engineering mechanisms to optimize performance. Additionally, the verifiability of user-side results is not fully addressed in this work and should be considered in future designs.

\Acknowledgements{This study was funded by the National Natural Science Foundation of China (No. 62471020), the Beijing Natural Science Foundation (L251066)}

\Supplements{Appendix A.}


\bibliographystyle{scis}
\bibliography{scis_paper}

\begin{thebibliography}{10}
\providecommand{\url}[1]{\texttt{#1}}
\providecommand{\urlprefix}{URL }
\providecommand{\bibinfo}[2]{#2}

\bibitem{monz2016realization}
\bibinfo{author}{Monz T}, \bibinfo{author}{Nigg D}, \bibinfo{author}{Martinez E~A}, et~al.
\newblock \bibinfo{title}{Realization of a scalable shor algorithm}.
\newblock \bibinfo{journal}{Science}, \bibinfo{year}{2016}, \bibinfo{volume}{351}: \bibinfo{pages}{1068--1070}

\bibitem{cao2018potential}
\bibinfo{author}{Cao Y}, \bibinfo{author}{Romero J}, \bibinfo{author}{Aspuru-Guzik A}.
\newblock \bibinfo{title}{Potential of quantum computing for drug discovery}.
\newblock \bibinfo{journal}{IBM Journal of Research and Development}, \bibinfo{year}{2018}, \bibinfo{volume}{62}: \bibinfo{pages}{6--1}

\bibitem{bauer2020quantum}
\bibinfo{author}{Bauer B}, \bibinfo{author}{Bravyi S}, \bibinfo{author}{Motta M}, et~al.
\newblock \bibinfo{title}{Quantum algorithms for quantum chemistry and quantum materials science}.
\newblock \bibinfo{journal}{Chemical reviews}, \bibinfo{year}{2020}, \bibinfo{volume}{120}: \bibinfo{pages}{12685--12717}

\bibitem{zou2025qpanda3}
\bibinfo{author}{Zou T}, \bibinfo{author}{Fang Y}, \bibinfo{author}{Wang J}, et~al.
\newblock \bibinfo{title}{Qpanda3: A high-performance software-hardware collaborative framework for large-scale quantum-classical computing integration}.
\newblock \bibinfo{journal}{arXiv preprint arXiv:2504.02455}, \bibinfo{year}{2025}

\bibitem{chow2021ibm}
\bibinfo{author}{Chow J}, \bibinfo{author}{Dial O}, \bibinfo{author}{Gambetta J}.
\newblock \bibinfo{title}{Ibm quantum breaks the 100-qubit processor barrier}.
\newblock \bibinfo{journal}{IBM Research Blog}, \bibinfo{year}{2021}, \bibinfo{volume}{2}

\bibitem{prateek2023quantum}
\bibinfo{author}{Prateek K}, \bibinfo{author}{Maity S}.
\newblock \bibinfo{title}{Quantum programming on azure quantum—an open source tool for quantum developers}.
\newblock In: \bibinfo{booktitle}{Quantum Computing: A Shift from Bits to Qubits}, \bibinfo{year}{2023}. \bibinfo{pages}{283--309}

\bibitem{ash2020analysis}
\bibinfo{author}{Ash-Saki A}, \bibinfo{author}{Alam M}, \bibinfo{author}{Ghosh S}.
\newblock \bibinfo{title}{Analysis of crosstalk in nisq devices and security implications in multi-programming regime}.
\newblock In: \bibinfo{booktitle}{Proceedings of the ACM/IEEE International Symposium on Low Power Electronics and Design}, \bibinfo{year}{2020}.
\newblock \bibinfo{pages}{25--30}

\bibitem{das2023trojannet}
\bibinfo{author}{Das S}, \bibinfo{author}{Ghosh S}.
\newblock \bibinfo{title}{Trojannet: Detecting trojans in quantum circuits using machine learning}.
\newblock \bibinfo{journal}{arXiv preprint arXiv:2306.16701}, \bibinfo{year}{2023}

\bibitem{das2024trojan}
\bibinfo{author}{Das S}, \bibinfo{author}{Ghosh S}.
\newblock \bibinfo{title}{Trojan attacks on variational quantum circuits and countermeasures}.
\newblock In: \bibinfo{booktitle}{2024 25th International Symposium on Quality Electronic Design (ISQED)}, \bibinfo{year}{2024}.
\newblock \bibinfo{pages}{1--8}

\bibitem{roy2024hardware}
\bibinfo{author}{Roy R}, \bibinfo{author}{Das S}, \bibinfo{author}{Ghosh S}.
\newblock \bibinfo{title}{Hardware trojans in quantum circuits, their impacts, and defense}.
\newblock In: \bibinfo{booktitle}{2024 25th International Symposium on Quality Electronic Design (ISQED)}, \bibinfo{year}{2024}.
\newblock \bibinfo{pages}{1--8}

\bibitem{john2025quantum}
\bibinfo{author}{John J}, \bibinfo{author}{Golla L}, \bibinfo{author}{Wang Q}.
\newblock \bibinfo{title}{Quantum trojan insertion: Controlled activation for covert circuit manipulation}.
\newblock \bibinfo{journal}{arXiv preprint arXiv:2502.08880}, \bibinfo{year}{2025}

\bibitem{xu2023exploration}
\bibinfo{author}{Xu C}, \bibinfo{author}{Erata F}, \bibinfo{author}{Szefer J}.
\newblock \bibinfo{title}{Exploration of power side-channel vulnerabilities in quantum computer controllers}.
\newblock In: \bibinfo{booktitle}{Proceedings of the 2023 ACM SIGSAC Conference on Computer and Communications Security}, \bibinfo{year}{2023}.
\newblock \bibinfo{pages}{579--593}

\bibitem{trochatos2023hardware}
\bibinfo{author}{Trochatos T}, \bibinfo{author}{Xu C}, \bibinfo{author}{Deshpande S}, et~al.
\newblock \bibinfo{title}{Hardware architecture for a quantum computer trusted execution environment}.
\newblock \bibinfo{journal}{arXiv preprint arXiv:2308.03897}, \bibinfo{year}{2023}

\bibitem{yang2024multi}
\bibinfo{author}{Yang M}, \bibinfo{author}{Guo X}, \bibinfo{author}{Jiang L}.
\newblock \bibinfo{title}{Multi-stage watermarking for quantum circuits}.
\newblock In: \bibinfo{booktitle}{2024 IEEE International Conference on Quantum Computing and Engineering (QCE)}, \bibinfo{year}{2024}, volume~\bibinfo{volume}{1}.
\newblock \bibinfo{pages}{796--804}

\bibitem{aboy2022mapping}
\bibinfo{author}{Aboy M}, \bibinfo{author}{Minssen T}, \bibinfo{author}{Kop M}.
\newblock \bibinfo{title}{Mapping the patent landscape of quantum technologies: patenting trends, innovation and policy implications}.
\newblock \bibinfo{journal}{IIC-International Review of Intellectual Property and Competition Law}, \bibinfo{year}{2022}, \bibinfo{volume}{53}: \bibinfo{pages}{853--882}

\bibitem{suresh2021short}
\bibinfo{author}{Suresh A}, \bibinfo{author}{Saki A~A}, \bibinfo{author}{Alam M}, et~al.
\newblock \bibinfo{title}{Short paper: A quantum circuit obfuscation methodology for security and privacy}.
\newblock In: \bibinfo{booktitle}{Proceedings of the 10th International Workshop on Hardware and Architectural Support for Security and Privacy}, \bibinfo{year}{2021}.
\newblock \bibinfo{pages}{1--5}

\bibitem{das2023randomized}
\bibinfo{author}{Das S}, \bibinfo{author}{Ghosh S}.
\newblock \bibinfo{title}{Randomized reversible gate-based obfuscation for secured compilation of quantum circuit}.
\newblock \bibinfo{journal}{arXiv preprint arXiv:2305.01133}, \bibinfo{year}{2023}

\bibitem{naz2023reversible}
\bibinfo{author}{Naz S~F}, \bibinfo{author}{Shah A~P}.
\newblock \bibinfo{title}{Reversible gates: A paradigm shift in computing}.
\newblock \bibinfo{journal}{IEEE Open Journal of Circuits and Systems}, \bibinfo{year}{2023}, \bibinfo{volume}{4}: \bibinfo{pages}{241--257}

\bibitem{saki2021split}
\bibinfo{author}{Saki A~A}, \bibinfo{author}{Suresh A}, \bibinfo{author}{Topaloglu R~O}, et~al.
\newblock \bibinfo{title}{Split compilation for security of quantum circuits}.
\newblock In: \bibinfo{booktitle}{2021 IEEE/ACM International Conference On Computer Aided Design (ICCAD)}, \bibinfo{year}{2021}.
\newblock \bibinfo{pages}{1--7}

\bibitem{upadhyay2022robust}
\bibinfo{author}{Upadhyay S}, \bibinfo{author}{Ghosh S}.
\newblock \bibinfo{title}{Robust and secure hybrid quantum-classical computation on untrusted cloud-based quantum hardware}.
\newblock In: \bibinfo{booktitle}{Proceedings of the 11th International Workshop on Hardware and Architectural Support for Security and Privacy}, \bibinfo{year}{2022}.
\newblock \bibinfo{pages}{45--52}

\bibitem{wang2025tetrislock}
\bibinfo{author}{Wang Q}, \bibinfo{author}{John J}, \bibinfo{author}{Dong B}, et~al.
\newblock \bibinfo{title}{Tetrislock: Quantum circuit split compilation with interlocking patterns}.
\newblock \bibinfo{journal}{arXiv preprint arXiv:2503.11982}, \bibinfo{year}{2025}

\bibitem{patel2023toward}
\bibinfo{author}{Patel T}, \bibinfo{author}{Silver D}, \bibinfo{author}{Ranjan A}, et~al.
\newblock \bibinfo{title}{Toward privacy in quantum program execution on untrusted quantum cloud computing machines for business-sensitive quantum needs}.
\newblock \bibinfo{journal}{arXiv preprint arXiv:2307.16799}, \bibinfo{year}{2023}

\bibitem{topaloglu2023quantum}
\bibinfo{author}{Topaloglu R~O}.
\newblock \bibinfo{title}{Quantum logic locking for security}.
\newblock \bibinfo{journal}{J}, \bibinfo{year}{2023}, \bibinfo{volume}{6}: \bibinfo{pages}{411--420}

\bibitem{liu2025loq}
\bibinfo{author}{Liu Y}, \bibinfo{author}{John J}, \bibinfo{author}{Wang Q}.
\newblock \bibinfo{title}{E-loq: Enhanced locking for quantum circuit ip protection}.
\newblock In: \bibinfo{booktitle}{2025 IEEE International Symposium on Hardware Oriented Security and Trust (HOST)}, \bibinfo{year}{2025}.
\newblock \bibinfo{pages}{67--77}

\bibitem{rehman2025opaque}
\bibinfo{author}{Rehman A}, \bibinfo{author}{Langford V}, \bibinfo{author}{John J}, et~al.
\newblock \bibinfo{title}{Opaque: Obfuscating phase in quantum circuit compilation for efficient ip protection}.
\newblock In: \bibinfo{booktitle}{2025 26th International Symposium on Quality Electronic Design (ISQED)}, \bibinfo{year}{2025}.
\newblock \bibinfo{pages}{1--6}

\bibitem{raj2025quantum}
\bibinfo{author}{Raj A}, \bibinfo{author}{Balachandran V}.
\newblock \bibinfo{title}{Quantum opacity, classical clarity: A hybrid approach to quantum circuit obfuscation}.
\newblock \bibinfo{journal}{arXiv preprint arXiv:2505.13848}, \bibinfo{year}{2025}

\bibitem{golec2024quantum}
\bibinfo{author}{Golec M}, \bibinfo{author}{Hatay E~S}, \bibinfo{author}{Golec M}, et~al.
\newblock \bibinfo{title}{Quantum cloud computing: Trends and challenges}.
\newblock \bibinfo{journal}{Journal of Economy and Technology}, \bibinfo{year}{2024}, \bibinfo{volume}{2}: \bibinfo{pages}{190--199}

\bibitem{broadbent2021constructions}
\bibinfo{author}{Broadbent A}, \bibinfo{author}{Kazmi R~A}.
\newblock \bibinfo{title}{Constructions for quantum indistinguishability obfuscation}.
\newblock In: \bibinfo{booktitle}{International Conference on Cryptology and Information Security in Latin America}, \bibinfo{year}{2021}.
\newblock \bibinfo{pages}{24--43}

\bibitem{bartusek2021indistinguishability}
\bibinfo{author}{Bartusek J}, \bibinfo{author}{Malavolta G}.
\newblock \bibinfo{title}{Indistinguishability obfuscation of null quantum circuits and applications}.
\newblock \bibinfo{journal}{arXiv preprint arXiv:2106.06094}, \bibinfo{year}{2021}

\bibitem{coladangelo2024use}
\bibinfo{author}{Coladangelo A}, \bibinfo{author}{Gunn S}.
\newblock \bibinfo{title}{How to use quantum indistinguishability obfuscation}.
\newblock In: \bibinfo{booktitle}{Proceedings of the 56th Annual ACM Symposium on Theory of Computing}, \bibinfo{year}{2024}.
\newblock \bibinfo{pages}{1003--1008}

\bibitem{shang2019obfuscatability}
\bibinfo{author}{Shang T}, \bibinfo{author}{Chen R~y~l}, \bibinfo{author}{Liu J~w}.
\newblock \bibinfo{title}{On the obfuscatability of quantum point functions}.
\newblock \bibinfo{journal}{Quantum Information Processing}, \bibinfo{year}{2019}, \bibinfo{volume}{18}: \bibinfo{pages}{55}

\bibitem{pan2023universal}
\bibinfo{author}{Pan C}, \bibinfo{author}{Shang T}, \bibinfo{author}{Zhang Y}.
\newblock \bibinfo{title}{Universal quantum obfuscation for quantum non-linear functions}.
\newblock \bibinfo{journal}{Frontiers in Physics}, \bibinfo{year}{2023}, \bibinfo{volume}{10}: \bibinfo{pages}{1048832}

\bibitem{boykin2003optimal}
\bibinfo{author}{Boykin P~O}, \bibinfo{author}{Roychowdhury V}.
\newblock \bibinfo{title}{Optimal encryption of quantum bits}.
\newblock \bibinfo{journal}{Physical review A}, \bibinfo{year}{2003}, \bibinfo{volume}{67}: \bibinfo{pages}{042317}

\bibitem{liang2013symmetric}
\bibinfo{author}{Liang M}.
\newblock \bibinfo{title}{Symmetric quantum fully homomorphic encryption with perfect security}.
\newblock \bibinfo{journal}{Quantum information processing}, \bibinfo{year}{2013}, \bibinfo{volume}{12}: \bibinfo{pages}{3675--3687}

\bibitem{aaronson2017ten}
\bibinfo{author}{S A}.
\newblock \bibinfo{title}{Ten semi-grand challenges for quantum computing theory}.
\newblock \bibinfo{journal}{URL: https://www. scottaaronson. com/writings/qchallenge. html}, \bibinfo{year}{2005}

\bibitem{alagic2016quantum}
\bibinfo{author}{Alagic G}, \bibinfo{author}{Fefferman B}.
\newblock \bibinfo{title}{On quantum obfuscation}.
\newblock \bibinfo{journal}{arXiv preprint arXiv:1602.01771}, \bibinfo{year}{2016}

\bibitem{zhang2022instantiation}
\bibinfo{author}{Zhang Y}, \bibinfo{author}{Shang T}, \bibinfo{author}{Chen R}, et~al.
\newblock \bibinfo{title}{Instantiation of quantum point obfuscation}.
\newblock \bibinfo{journal}{Quantum Information Processing}, \bibinfo{year}{2022}, \bibinfo{volume}{21}: \bibinfo{pages}{29}

\bibitem{jiang2023quantum}
\bibinfo{author}{Jiang Y}, \bibinfo{author}{Shang T}, \bibinfo{author}{Tang Y}, et~al.
\newblock \bibinfo{title}{Quantum obfuscation of generalized quantum power functions with coefficient}.
\newblock \bibinfo{journal}{Entropy}, \bibinfo{year}{2023}, \bibinfo{volume}{25}: \bibinfo{pages}{1524}

\bibitem{bartusek2024quantum}
\bibinfo{author}{Bartusek J}, \bibinfo{author}{Brakerski Z}, \bibinfo{author}{Vaikuntanathan V}.
\newblock \bibinfo{title}{Quantum state obfuscation from classical oracles}.
\newblock In: \bibinfo{booktitle}{Proceedings of the 56th Annual ACM Symposium on Theory of Computing}, \bibinfo{year}{2024}.
\newblock \bibinfo{pages}{1009--1017}

\bibitem{bernstein1993quantum}
\bibinfo{author}{Bernstein E}, \bibinfo{author}{Vazirani U}.
\newblock \bibinfo{title}{Quantum complexity theory}.
\newblock In: \bibinfo{booktitle}{Proceedings of the twenty-fifth annual ACM symposium on Theory of computing}, \bibinfo{year}{1993}.
\newblock \bibinfo{pages}{11--20}

\bibitem{jain2022indistinguishability}
\bibinfo{author}{Jain A}, \bibinfo{author}{Jin Z}.
\newblock \bibinfo{title}{Indistinguishability obfuscation via mathematical proofs of equivalence}.
\newblock In: \bibinfo{booktitle}{2022 IEEE 63rd Annual Symposium on Foundations of Computer Science (FOCS)}, \bibinfo{year}{2022}.
\newblock \bibinfo{pages}{1023--1034}

\bibitem{shang2023two}
\bibinfo{author}{Shang T}, \bibinfo{author}{Wang S}, \bibinfo{author}{Jiang Y}, et~al.
\newblock \bibinfo{title}{Two-round quantum homomorphic encryption scheme based on matrix decomposition: T. shang et al.}
\newblock \bibinfo{journal}{Quantum Information Processing}, \bibinfo{year}{2023}, \bibinfo{volume}{22}: \bibinfo{pages}{422}

\bibitem{bravyi2016improved}
\bibinfo{author}{Bravyi S}, \bibinfo{author}{Gosset D}.
\newblock \bibinfo{title}{Improved classical simulation of quantum circuits dominated by clifford gates}.
\newblock \bibinfo{journal}{Physical review letters}, \bibinfo{year}{2016}, \bibinfo{volume}{116}: \bibinfo{pages}{250501}

\bibitem{gottesman1998heisenberg}
\bibinfo{author}{Gottesman D}.
\newblock \bibinfo{title}{The heisenberg representation of quantum computers}.
\newblock \bibinfo{journal}{arXiv preprint quant-ph/9807006}, \bibinfo{year}{1998}

\bibitem{belavkin1994nondemolition}
\bibinfo{author}{Belavkin V~P}.
\newblock \bibinfo{title}{Nondemolition principle of quantum measurement theory}.
\newblock \bibinfo{journal}{Foundations of Physics}, \bibinfo{year}{1994}, \bibinfo{volume}{24}: \bibinfo{pages}{685--714}

\bibitem{zhang2024quantum}
\bibinfo{author}{Zhang Y}, \bibinfo{author}{Shang T}, \bibinfo{author}{Zhang K}, et~al.
\newblock \bibinfo{title}{Quantum indistinguishable obfuscation via quantum circuit equivalence}.
\newblock \bibinfo{journal}{arXiv preprint arXiv:2411.12297}, \bibinfo{year}{2024}

\bibitem{amy2018towards}
\bibinfo{author}{Amy M}.
\newblock \bibinfo{title}{Towards large-scale functional verification of universal quantum circuits}.
\newblock \bibinfo{journal}{arXiv preprint arXiv:1805.06908}, \bibinfo{year}{2018}

\bibitem{das2021adapt}
\bibinfo{author}{Das P}, \bibinfo{author}{Tannu S}, \bibinfo{author}{Dangwal S}, et~al.
\newblock \bibinfo{title}{Adapt: Mitigating idling errors in qubits via adaptive dynamical decoupling}.
\newblock In: \bibinfo{booktitle}{MICRO-54: 54th Annual IEEE/ACM International Symposium on Microarchitecture}, \bibinfo{year}{2021}.
\newblock \bibinfo{pages}{950--962}

\bibitem{motwani1996randomized}
\bibinfo{author}{Motwani R}, \bibinfo{author}{Raghavan P}.
\newblock \bibinfo{title}{Randomized algorithms}.
\newblock \bibinfo{journal}{ACM Computing Surveys (CSUR)}, \bibinfo{year}{1996}, \bibinfo{volume}{28}: \bibinfo{pages}{33--37}

\bibitem{xu2023synthesizing}
\bibinfo{author}{Xu A}, \bibinfo{author}{Molavi A}, \bibinfo{author}{Pick L}, et~al.
\newblock \bibinfo{title}{Synthesizing quantum-circuit optimizers}.
\newblock \bibinfo{journal}{Proceedings of the ACM on Programming Languages}, \bibinfo{year}{2023}, \bibinfo{volume}{7}: \bibinfo{pages}{835--859}

\bibitem{burgholzer2020advanced}
\bibinfo{author}{Burgholzer L}, \bibinfo{author}{Wille R}.
\newblock \bibinfo{title}{Advanced equivalence checking for quantum circuits}.
\newblock \bibinfo{journal}{IEEE Transactions on Computer-Aided Design of Integrated Circuits and Systems}, \bibinfo{year}{2020}, \bibinfo{volume}{40}: \bibinfo{pages}{1810--1824}

\bibitem{peham2022equivalence}
\bibinfo{author}{Peham T}, \bibinfo{author}{Burgholzer L}, \bibinfo{author}{Wille R}.
\newblock \bibinfo{title}{Equivalence checking of quantum circuits with the zx-calculus}.
\newblock \bibinfo{journal}{IEEE Journal on Emerging and Selected Topics in Circuits and Systems}, \bibinfo{year}{2022}, \bibinfo{volume}{12}: \bibinfo{pages}{662--675}

\bibitem{wille2008revlib}
\bibinfo{author}{Wille R}, \bibinfo{author}{Gro{\ss}e D}, \bibinfo{author}{Teuber L}, et~al.
\newblock \bibinfo{title}{Revlib: An online resource for reversible functions and reversible circuits}.
\newblock In: \bibinfo{booktitle}{38th International Symposium on Multiple Valued Logic (ismvl 2008)}, \bibinfo{year}{2008}.
\newblock \bibinfo{pages}{220--225}

\bibitem{jin2024quafu}
\bibinfo{author}{Jin Y~X}, \bibinfo{author}{Xu H~Z}, \bibinfo{author}{Wang Z~A}, et~al.
\newblock \bibinfo{title}{Quafu-rl: The cloud quantum computers based quantum reinforcement learning}.
\newblock \bibinfo{journal}{Chinese Physics B}, \bibinfo{year}{2024}, \bibinfo{volume}{33}: \bibinfo{pages}{050301}

\bibitem{bartake2025obfusqate}
\bibinfo{author}{Bartake N}, \bibinfo{author}{Jie S~T~Z}, \bibinfo{author}{Jiawen C~W}, et~al.
\newblock \bibinfo{title}{Obfusqate: Unveiling the first quantum program obfuscation framework}.
\newblock \bibinfo{journal}{arXiv preprint arXiv:2503.23785}, \bibinfo{year}{2025}

\end{thebibliography}







\end{document}


\ArticleType{Supplementary File}

\title{Encrypted-state quantum compilation scheme based on quantum circuit obfuscation for quantum cloud platforms}{}

\author[1]{Chenyi Zhang}{}
\author[1]{Tao Shang}{{shangtao@buaa.edu.cn}}
\author[2]{Xueyi Guo}{}


\address[1]{School of Cyber Science and Technology, Beihang University, Beijing 100083, China}
\address[2]{Beijing Academy of Quantum Information Sciences, Beijing 100193, China}

\maketitle


\begin{appendix}

\section{$T/T^\dagger$ gate replacement for leaked key}
In Appendix A, we provide a theoretical explanation of why applying the $U$ gate substitution rule to replace $T/T^\dagger$ gates in QCOO's key update process may lead to user key leakage. Taking the $T$ gate as an example, its ZYZ decomposition is shown in Equation \ref{eq_ZYZ}.
\begin{align}\label{eq_ZYZ}
    \nonumber T=&e^{i\alpha}R_z(\beta)R_y(\gamma)R_z(\delta)\\
    =&e^{i\alpha}\left[\begin{array}{cc}e^{-i\beta/2}&0 \\0&e^{i\beta/2}\end{array}\right]\left[\begin{array}{cc}\mathrm{cos}\frac{\gamma}{2}&\mathrm{-sin}\frac{\gamma}{2} \\\mathrm{sin}\frac{\gamma}{2}&\mathrm{cos}\frac{\gamma}{2}\end{array}\right]\left[\begin{array}{cc}e^{-i\delta/2}&0 \\0&e^{i\delta/2}\end{array}\right]\\\nonumber
    =&e^{i\alpha}\left[\begin{array}{cc}e^{-i(\beta+\delta)/2}\mathrm{cos}\frac{\gamma}{2}&-e^{-i(\beta-\delta)/2}\mathrm{sin}\frac{\gamma}{2} \\e^{i(\beta-\delta)/2}\mathrm{sin}\frac{\gamma}{2}&e^{i(\beta+\delta)/2}\mathrm{cos}\frac{\gamma}{2}\end{array}\right]
\end{align}

By comparing the form of the T gate, we can obtain that $\gamma=0,\alpha-\frac{\beta+\delta}{2}=0,\alpha+\frac{\beta+\delta}{2}=\pi/4$. The T gate can be decomposed into $T=U(\pi/8,\beta,0,\delta)$, where $\beta+\delta=\pi/4$. According to the key and gate substitution rules, $T$ gates in the quantum circuit are replaced with $T'$, as shown in Equation \ref{eq_t'}.

\begin{align}\label{eq_t'}
\nonumber T=&U(\pi/8,(-1)^a\beta,0,(-1)^a\delta)\\
=&e^{i\alpha}\left[\begin{array}{cc}e^{i(-1)^{a+1}(\beta+\delta)/2}&0\\0&e^{i(-1)^a\beta+\delta)/2}\end{array}\right]\\\nonumber
=&\left[\begin{array}{cc}e^{i\pi/8(1+(-1)^{a+1})}&0\\0&e^{i\pi/8(1+(-1)^a}\end{array}\right]
\end{align}

The computing party can obtain the replaced gate $T'$ and decompose it, then there is $T'=U(\pi/8,\beta',\gamma',\delta')$. Only $T$ gate in the quantum circuit is replaced, so there must be Equation \ref{eq_t'depos} holding.
\begin{align}\label{eq_t'depos}
   T'=U(\pi/8,\beta',\gamma',\delta')=U(\pi/8,(-1)^a\beta,0,(-1)^a\delta)
\end{align}

Therefore, $\beta'+\delta'=(-1)^a\beta+(-1)^a\delta=(-1)^a(\beta+\delta)=(-1)^a\pi/4$. According to the decomposition results of $T'$, if $\beta'+\delta'=\pi/4$, then $a=0$. If $\beta'+\delta'=-\pi/4$, then $a=1$. The same applies to $T^\dagger$. Since the computing party can distinguish between $T$ gates and $T^\dagger$ gates through $\alpha$. Hence, the computing party only needs to extract the parameters of $T$ gates and $T^\dagger$ gates, and then infer the value of the key through consistency. The computing party can obtain the key by comparing the quantum gates before and after the gate replacement.

\section{Polynomial equivalence detection}
The correctness of QCSO relies on verifying the obfuscated quantum circuit using the Probabilistic
Testing Distinguisher (PTD). In Appendix B, we further explain the idea behind PTD and provide a proof of its validity. PTD draws inspiration from polynomial identity testing \cite{xu2023synthesizing}, a method widely used in integrated circuit verification and semantic optimization. It reduces the indistinguishability verification of quantum circuits to a subpath sum equivalence problem, and performs probabilistic testing on the phase polynomials of these subpaths. 

Assume the quantum circuit is $C$, and $m$ is the number of qubit outputs of $C$. The phase polynomial $\phi\in D[x,y] $ is a linear polynomial in the input variable $x$ and the path variables $y=y_1y_2\dots y_m$. The phase polynomial encodes the relative phase accumulated along each computational path. Given $\phi_1$ and $\phi_2$,$\phi_1,\phi_2\in D_M[x,y]$ defined over the same set of $n$ variables, the procedure inserts $\ell$ polynomials into $\phi_1$ and checks whether for all variables $\upsilon\in\mathbb{C}^n$, there is $\phi_1(\upsilon)=\phi_2(\upsilon)$, concisely represented as $\phi_1=\phi_2$. $d$ represents the maximum degree of the two polynomials. The correctness of PTD is shown in Theorem \ref{t_ptd}

\theorem [The correctness of PTD]{
 PTD can verify the equivalence of $\phi_1$ and $\phi_2$ with a high probability $1-\ell^2d/|R|$.}\label{t_ptd}

\begin{proof}
Suppose we insert $\ell$ mutually non-equivalent polynomials into $\phi_1$, all defined over the same set of $n$ variables. This defines an implicit randomized computation consisting of the following steps:
 \begin{enumerate}
     \item Select a finite subset $R\in\mathbb{C}$ of complex numbers.
    \item Sample $n$ independent values from $R$, $\upsilon_1,\dots,\upsilon_n$.
    \item For each pair $\phi_i$ and $\phi_j$, check whether $\phi_i(\upsilon)=\phi_j(\upsilon),i\neq j,\quad i,j\in\{1,\dots,\ell\}$.
 \end{enumerate}
 
According to the Schwartz–Zippel lemma \cite{motwani1996randomized} (which provides a worst-case bound), the algorithm returns True if $\phi_1=\phi_2$. This is because $\forall \upsilon,\phi_1(\upsilon)=\phi_2(\upsilon)$. If $\phi_1\neq\phi_2$, then the probability of returning True is at most $d/|R|$. Otherwise, the probability of returning False is at least $1-d/|R|$. For $\phi_1\neq\phi_2$, there is a small chance that the answer is incorrect. Therefore, by the union bound, for any pair of polynomials $\phi_i,\phi_j$,  if $\phi_i(\upsilon)=\phi_j(\upsilon)$, the test returns True. If $\phi_i(\upsilon)\neq\phi_j(\upsilon)$, the probability of returning True satisfies Equation \ref{eq_pr}.
\begin{align}\label{eq_pr}
    \mathrm{Pr}\big[\exists i\neq j, \phi_i(\upsilon)=\phi_j(\upsilon)\big]\leq \sum\limits_{i,j\in[1,\ell],i\neq j}   \mathrm{Pr}\big[ \phi_i(\upsilon)=\phi_j(\upsilon)\big] \leq \ell^2d/|R|
\end{align}

Otherwise, the algorithm returns False with probability at least $1-\ell^2d/|R|$. The proof is complete.
\end{proof}

For example, suppose $\phi_1$ and $\phi_2$  are two non-equivalent polynomials of degree 10. If the set $R$ is chosen as 64-bit integers ($|R|=2^{64}\approx 10^{20}$), the probability of a false positive is around $10^{-19}$. When $10^6$ non-equivalent phase polynomials, each of degree at most 10, are inserted into a new phase polynomial, and 64-bit integers are still used for $R$, the chance of mistakenly declaring two polynomials as equal increases to approximately $10^{-7}$. Although the error accumulates with more insertions when the polynomials are unequal, the number of inserted phase polynomials in practice is much smaller than $10^6$. This is due to hardware-level decoherence and circuit structure constraints in quantum computation. Therefore, the overall failure probability remains very low, thus ensuring the correctness of PTD.

\end{appendix}

\bibliographystyle{scis}
\bibliography{scis_paper}